\documentclass[pre,preprint,showpacs,superscriptaddress]{revtex4-1}
\usepackage{graphicx}
\usepackage{amsmath}
\usepackage{amssymb}

\begin{document}

\title{Particle in a box with a time-dependent $\delta$-function potential}
\author{Seung Ki Baek}
\email{seungki@pknu.ac.kr}
\affiliation{Department of Physics, Pukyong National University, Busan 48513,
Korea}
\author{Su Do Yi}
\affiliation{Department of Physics and Astronomy, Seoul National University,
Seoul 08826, Korea}
\author{Minjae Kim}
\affiliation{Department of Physics, Pukyong National University, Busan 48513,
Korea}

\begin{abstract}
In quantum information processing, one often considers inserting a barrier into
a box containing a particle to generate one bit of Shannon entropy.
We formulate this problem as a one-dimensional Schr\"{o}dinger equation with a
time-dependent $\delta$-function potential. It is a natural generalization of
the particle in a box, a canonical example of quantum mechanics, and we present
analytic and numerical investigations on this problem. After deriving
an exact Volterra-type integral equation, composed of an infinite sum of
modes, we show that approximate formulas
with the lowest-frequency modes correctly capture the qualitative behavior
of the wave function. If we take into account hundreds of modes, our numerical
calculation shows that the quantum adiabatic theorem actually gives a very good
approximation even if the barrier height diverges within finite time, as long
as it is sufficiently longer than the characteristic time scale of the particle.
In particular, if the barrier is slowly inserted at an asymmetric position, the
particle is localized by the insertion itself, in accordance with a prediction
of the adiabatic theorem. On the other hand, when the barrier is inserted
quickly, the wave function becomes rugged after the insertion because of the
energy transfer to the particle. Regardless of the position of the barrier,
the fast insertion leaves the particle unlocalized so that we can obtain
meaningful information by a which-side measurement. Our numerical procedure
provides a precise way to calculate the wave function throughout the process,
from which one can estimate the amount of this information for an arbitrary
insertion protocol.
\end{abstract}

\maketitle

\section{Introduction}

The ``particle in a box'' problem describes a localized particle in a deep
potential well. It is one of the most basic problems in textbooks of quantum
mechanics, yet relevant in many physical situations. In particular, the
mathematical simplicity makes it a useful starting point to study many different
phenomena, such as quantum dots~\cite{alivisatos}, ideal gases~\cite{swendsen},
and energy bands in a periodic crystal lattice~\cite{kittel}. An important
application of the particle in a box problem is the Szilard
engine~\cite{szilard,zurek,swkim}, which has drawn attention in the context of
information
thermodynamics~\cite{sagawa08,jacobs2,toyabe,sagawa10,*sagawa12,*sagawa12b,*sagawa13,jacobs1,horowitz,um}.
The original Szilard engine is illustrated as a classical particle in a box,
in the middle of which an impenetrable barrier is inserted to
create uncertainty of one bit~\cite{[{The performance of the engine was
considered in }] [{ for the case of a penetrable barrier.}]park}.
The uncertainty is resolved by measuring the position of the
particle, and the resulting information can be used to extract work from a
single thermal reservoir. After the work extraction, the barrier is removed from
the box, which completes one cycle.
Since the criticism on thermodynamics of the single-particle gas~\cite{jauch},
however, it has been argued that the Szilard engine requires
quantum-mechanical treatment~\cite{zurek}.
Along this line, for example, one can model the working substance of the engine
as a particle governed by the Schr{\"o}dinger equation with a time-dependent
potential barrier.

If we are to take into account the time dependence,
one of the easiest ways would be to consider an infinitely slow protocol for
changing the barrier height in an isolated system, to which the adiabatic
theorem is applicable~\cite{messiah}. The occupation probabilities then remain
unchanged while the energy levels are shifted. Another extreme case is to
insert an infinitely high barrier all of sudden at time $t=0$, which has been
claimed to generate an unusual quantum state with a fractal
wave function~\cite{bender}.
These two protocols bring the system out of equilibrium at the end of
the process, at which isothermal expansion gets started. To avoid abrupt
transitions at this moment, one may alternatively consider an isothermal
process, throughout which the system evolves quasistatically in contact with a
thermal reservoir~\cite{swkim,kim2}.
If the reservoir has very low temperature, however, this protocol runs into
a trouble: The thermal contact always brings the system to the
ground state. If the ground state is nondegenerate, it means that the system
has no uncertainty so that no work can be extracted by the engine~\cite{kim2}.
It happens when the barrier is off the middle of the box, which should always be
the case in experiments.

Motivated by the quantum Szilard engine,
we in this work calculate the time evolution of a quantum particle in an
isolated box when the $\delta$-function barrier is inserted with a finite speed.
This model deserves attention in its own right as a generalization of
the particle in a box problem, because the Schr{\"o}dinger equation is not
explicitly solvable in the presence of a time-dependent potential except a few
special cases~\cite{lewis,ji,truscott,feng,park02}.
It is also of experimental relevance in the context of splitting matter waves
such as Bose-Einstein
condensates~\cite{pezze,hohenester,*grond09a,*grond09b,masuda08,*masuda09,torrontegui}.
This topic has been investigated over the past decade since the experimental
realization of a stable double-well trap, which can be used for implementing
an atom interferometer~\cite{shin} or a Josephson junction~\cite{albiez}.
Researchers have mostly considered deforming a harmonic trap to split a
wave function, but it would also be experimentally feasible to localize a
wave function inside a deep square potential well and split it with a sharp
laser beam.
The solution for a finite-speed protocol will be generally useful in the
sense that it describes an experimentally accessible situation, whereas
infinitely slow or fast protocols are only approximate to reality.
Furthermore, if we are interested in the power of an engine in
evaluating the performance~\cite{broeck,esposito,kawai}, we must definitely
consider a finite-speed protocol, and this work can provide a starting point
in this direction as well.

A strategy to solve the Schr\"{o}dinger equation with a time-dependent
$\delta$-function barrier is to convert it to a Volterra-type integral
equation,
as suggested by Ref.~\onlinecite{campbell}. We will demonstrate how this
equation can be studied in a systematic way, in combination with analytic and
numerical calculations.
This work is organized as follows. In the next section, we derive the
Volterra-type integral equation from the Schr\"{o}dinger equation.
Some approximate expressions will also be given there.
In Sec.~\ref{sec:numeric}, we present a numerical procedure to solve the
integral equations and discuss the results. We then conclude this
work in Sec.~\ref{sec:conclusion}.

\section{Analysis}

\subsection{Formal solution}

The Schr\"{o}dinger equation is written in dimensionless units as
\begin{equation}
-\psi_{xx} + 2c(t) \delta(x-x_0) \psi = i\psi_t,
\label{eq:sch}
\end{equation}
where the subscripts mean partial derivatives.
The strength of the $\delta$-function potential is controlled by $c(t)$, which
is called a protocol in this work.
Let us choose our initial condition at $t=0$ as the $n$th
eigenstate $\phi_n(x)$ in the absence of the $\delta$-function potential:
\begin{equation}
\psi(x,0) = \phi_n(x) = \left\{
\begin{array}{ll}
L^{-1/2} \cos n k x & \mbox{if~}|x|<L \mbox{~and~$n$ is odd},\\
L^{-1/2} \sin n k x & \mbox{if~}|x|<L \mbox{~and~$n$ is even},\\
0 & \mbox{otherwise,}
\end{array}
\right.
\end{equation}
where $k \equiv \pi/(2L)$ and $n=1$ corresponds to the ground state.
The particle is confined in a box ranging from $x=-L$ to $x=+L$, so that the
boundary condition is given by $\psi(x,t) = 0$ at $x = \pm L$.
If $-L<x<x_0$ or $x_0<x<L$, we are
back to the free space described by $-\psi_{xx} = i\psi_t$. We restrict
ourselves to $t>0$ and take the Laplace transform $\mathcal{L}$ to obtain
\begin{equation}
\overline\psi_{xx} + is \overline\psi = i\psi(x,0),
\label{eq:lt}
\end{equation}
where $\overline\psi(x,s) \equiv \mathcal{L}\left[ \psi(x,t) \right] =
\int_0^\infty dt e^{-st} \psi(x,t)$ with $s>0$.
To solve the Schr\"{o}dinger equation with a time-dependent potential,
we transform it to an integral equation as suggested in
Ref.~\onlinecite{campbell}.
In Appendix~\ref{appendix:laplace}, we derive the following solution,
\begin{equation}
\overline\psi(x,s) = \frac{\phi_n(x)}{s+in^2k^2} + \mathcal{L}[c(t) \psi(x_0,t)]
F(x,s),
\label{eq:psixt}
\end{equation}
where
\begin{equation}
F(x,s) = \left\{
\begin{array}{ll}
\frac{\left( e^{2i\sqrt{is}L}-e^{2i\sqrt{is}x} \right)
\left( e^{2i\sqrt{is}(L+x_0)} - 1 \right)}
{i\sqrt{is} e^{i\sqrt{is}(x+x_0)} \left( e^{4i\sqrt{is}L} - 1 \right)}
& \mbox{for~} x_0 \le x < L,\\
\frac{\left( e^{2i\sqrt{is}L}-e^{2i\sqrt{is}x_0} \right)
\left( e^{2i\sqrt{is}(L+x)} - 1 \right)}
{i\sqrt{is} e^{i\sqrt{is}(x+x_0)} \left( e^{4i\sqrt{is}L} - 1 \right)}
& \mbox{for~} -L < x < x_0.
\end{array}
\right.
\label{eq:Fs}
\end{equation}
In Appendix~\ref{appendix:inverse}, we perform the inverse Laplace transform to
obtain
\begin{equation}
\psi(x,t) = \phi_n(x) e^{-in^2 k^2 t} + \sum_{\nu=1}^\infty
\int_0^t dt' c(t') \psi(x_0,t') f_\nu(x, t-t'),
\label{eq:volterra}
\end{equation}
where
\begin{equation}
f_\nu(x,t) = \left\{
\begin{array}{ll}
\frac{1}{2iL} 
e^{-i\nu^2 k^2 t - i\nu k(x+x_0)} \left[ (-1)^\nu - e^{2i\nu kx} \right] \left[
e^{2i \nu k(L+x_0)} -1 \right] & \mbox{for~} x_0 < x< L\\
\frac{1}{2iL} 
e^{-i\nu^2 k^2 t - i\nu k(x+x_0)} \left[ (-1)^\nu - e^{2i\nu kx_0} \right]\left[
e^{2i \nu k(L+x)} -1 \right] & \mbox{for~} -L < x< x_0.
\end{array} \right.
\end{equation}
Equation~\eqref{eq:volterra} has to be solved in two steps.
First, we solve it for $x=x_0$ to get $\psi(x_0,t)$.
The next step is to substitute $\psi(x_0,t)$ back into Eq.~\eqref{eq:volterra}
to obtain $\psi(x,t)$. Note that $\psi(x_0,t)$ contains essential information of
the full wave function in this formulation.

\subsection{Insertion at the origin}

To proceed, it is more convenient to deal with a specific situation.
Suppose that the system is initially in the ground state, i.e., $\psi(x,0) =
\phi_1(x)$.
If we insert the $\delta$-potential barrier at the origin by setting $x_0=0$,
Eq.~\eqref{eq:volterra} reduces to
\begin{equation}
\psi(x,t) = L^{-1/2} e^{-ik^2 t} \cos kx - 2iL^{-1} \sum_{\mu=0}^{\infty}
\cos[(2\mu+1)kx] \int_0^t dt' c(t') \psi(0,t') e^{-i (2\mu+1)^2 k^2 (t-t')}.
\label{eq:special}
\end{equation}
We may express Eq.~\eqref{eq:special} as the cosine series
\begin{equation}
\psi(x,t) = L^{-1/2} \sum_{\mu=0}^\infty \sigma_\mu(t) \cos [(2\mu+1)kx],
\label{eq:general}
\end{equation}
with $\sigma_\mu(t) \equiv e^{-ik^2 t} \delta_{\mu 0} - 2iL^{-1/2} \int_0^t dt'
c(t') \psi(0,t') e^{-i(2\mu+1)^2 k^2 (t-t')}$, where $\delta_{\alpha \beta}$
means the Kronecker $\delta$. The conservation of total probability implies that
\begin{equation}
\sum_{\mu=0}^\infty \left| \sigma_\mu(t) \right|^2 = 1,
\label{eq:normalization}
\end{equation}
because
$L^{-1} \int_{-L}^L dx \cos[(2\mu+1)kx] \cos[(2\mu'+1)kx] = \delta_{\mu \mu'}$
for integers $\mu$ and $\mu'$.
The total energy of the particle is obtained as
\begin{equation}
E(t) = \sum_{\mu=0}^\infty (2\mu+1)^2 k^2 \left| \sigma_\mu(t) \right|^2 +
2c(t) \left| \sum_{\mu=0}^\infty \sigma_\mu(t) \right|^2,
\label{eq:energy}
\end{equation}
where the first and second terms represent the kinetic and potential parts,
respectively.

\subsubsection{Single-mode approximation}
In our specific case described by Eq.~\eqref{eq:special},
let us define
\begin{equation}
g(x,t;\mu) \equiv 2iL^{-1} \cos[(2\mu+1)kx]
\int_0^t dt' c(t') \psi(0,t') e^{-i (2\mu+1)^2 k^2 (t-t')}
\end{equation}
to rewrite Eq.~\eqref{eq:special} as
\begin{equation}
\psi(x,t) = L^{-1/2} e^{-ik^2 t} \cos kx - \sum_{\mu=0}^{\infty} g(x,t;\mu).
\label{eq:withg}
\end{equation}
It is straightforward to obtain the following equality for
the time derivative of $g$:
\begin{equation}
g_t (x,t;\mu) = [-i (2\mu+1)^2 k^2] g(x,t;\mu) + 2iL^{-1}
c(t) \psi(0,t) \cos[(2\mu+1)kx].
\end{equation}
If we assume that only the lowest-frequency mode with $\mu = 0$ is dominant,
Eq.~\eqref{eq:withg} is simplified to
\begin{equation}
\psi(x,t) \approx L^{-1/2} e^{-ik^2 t} \cos kx - g(x,t;0).
\label{eq:lowest}
\end{equation}
Taking the time derivative, we find that
\begin{eqnarray}
\psi_t (x,t) &\approx&
L^{-1/2} (-ik^2) e^{-ik^2 t} \cos kx - (-ik^2)g(x,t;0) - 2iL^{-1}
c(t)\psi(0,t) \cos kx\nonumber\\
&\approx&
L^{-1/2} (-ik^2) e^{-ik^2 t} \cos kx\nonumber\\
&&- (-ik^2) [L^{-1/2} e^{-ik^2 t} \cos kx - \psi(x,t)] -
2iL^{-1} c(t)\psi(t) \cos kx\nonumber\\
&=& -ik^2 \psi(x,t) - 2iL^{-1} c(t) \psi(0,t) \cos kx,
\label{eq:single}
\end{eqnarray}
where we have used Eq.~\eqref{eq:lowest} to remove $g(x,t;0)$.
Rearranging the terms, we obtain the equation
\begin{equation}
\left( \frac{\partial}{\partial t} + ik^2 \right) \psi(x,t) \approx
-2iL^{-1} c(t) \psi(0,t) \cos kx,
\end{equation}
where the right-hand side (RHS) represents a `driving' term due to the
$\delta$-potential barrier.
If we regard $c(t) \psi(0,t)$ on the RHS as an external parameter,
we can write the formal solution of the first-order ordinary differential
equation (ODE)~\cite{boas}, which is identical to Eq.~\eqref{eq:lowest}.
As mentioned above, the evolution at the insertion point is most important in
determining the full wave function in our formulation. At $x=0$, we see from
Eq.~\eqref{eq:single} that
\begin{equation}
\psi_t(0,t) \approx -i[ k^2 + 2L^{-1} c(t)] \psi(0,t),
\end{equation}
for which the solution is obtained as
\begin{equation}
\psi(0,t) \approx \psi(0,0) \exp \left[-i \left(k^2 t + 2L^{-1} \int_0^t dt'
c(t') \right) \right].
\end{equation}
The lowest-mode approximation thus describes an effect of $c(t)$ on the
phase of $\psi(0,t)$ without altering the magnitude.

\subsubsection{Two-mode approximation}

\begin{figure}
\includegraphics[width=0.45\textwidth]{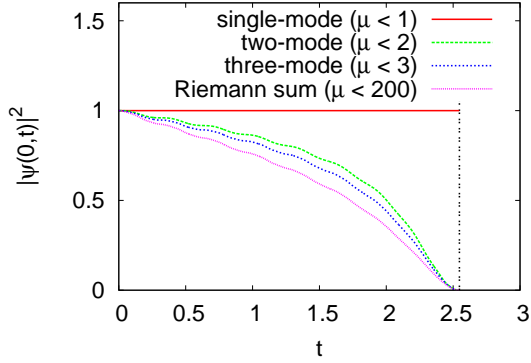}
\caption{(Color online) Probability density at $x=x_0=0$, when
the particle is assumed to be in the ground state at $t=0$.
The protocol is chosen as $c(t) = \tan (k^2t/4)$ with $L \equiv 1$. The
vertical line represents $t^\ast = 2\pi/k^2 = 8/\pi \approx 2.55$, where the
barrier height diverges to infinity. The single-mode approximation
[Eq.~\eqref{eq:single}] does not describe the amplitude change, but the two-mode
approximation is already plausible, and the result is further improved in
the three-mode approximation. The Riemann sum means the calculation explained in
Sec.~\ref{sec:numeric}. We use the second-order Runge-Kutta method~\cite{newman}
to integrate the ODEs resulting from the two- and
three-mode approximations such as Eq.~\eqref{eq:two_mode}.
}
\label{fig:two_mode}
\end{figure}

If we additionally take into account the next mode with $\mu=1$, we find the
following set of equations by taking time derivatives:
\begin{eqnarray}
\psi(x,t) &\approx& L^{-1/2} e^{-ik^2 t} \cos kx - g(x,t;0) -
g(x,t;1)\label{eq:formal2}\\
\psi_t(x,t) &\approx& L^{-1/2} (-ik^2) e^{-ik^2 t} \cos kx\nonumber\\
&&- [(-ik^2) g(x,t;0) + 2iL^{-1} c(t)\psi(0,t) \cos kx]\nonumber\\
&&- [(-9ik^2) g(x,t;1) + 2iL^{-1} c(t) \psi(0,t) \cos 3kx]\\
\psi_{tt}(x,t) &\approx& L^{-1/2} (-ik^2)^2 e^{-ik^2 t} \cos kx\nonumber\\
&&- (-ik^2) [(-ik^2) g(x,t;0) + 2iL^{-1} c(t) \psi(0,t) \cos kx]\nonumber\\
&& -2iL^{-1} c_t(t) \psi(0,t) \cos kx - 2iL^{-1} c(t) \psi_t(0,t) \cos
kx\nonumber\\
&&- (-9ik^2) [(-9ik^2) g(x,t;1) + 2iL^{-1} c(t) \psi(0,t) \cos 3kx]\nonumber\\
&& -2iL^{-1} c_t(t) \psi(0,t) \cos 3kx - 2iL^{-1} c(t) \psi_t(0,t) \cos
3kx.
\end{eqnarray}
By using the first two equations, one can remove $g(x,t;0)$ and $g(x,t;1)$ in
the last one to derive a second-order ODE for
$\psi(x,t)$:
\begin{eqnarray}
\left( \frac{\partial}{\partial t} + ik^2 \right) \left(
\frac{\partial}{\partial t} + 9ik^2 \right) \psi(x,t)
&\approx& -2iL^{-1}
\left( \frac{\partial}{\partial t} +ik^2 \right) [c(t)\psi(0,t)] \cos kx
\nonumber\\
&&-2iL^{-1} \left( \frac{\partial}{\partial t} +9ik^2 \right) [c(t)\psi(0,t)]
\cos 3kx.
\label{eq:two_mode_x}
\end{eqnarray}
The left-hand side (LHS) is related to the two lowest modes at $t=0$, whereas
the RHS represents the effect of the barrier interacting with the wave function
at $x=0$. Once again, the formal solution of Eq.~\eqref{eq:two_mode_x} for
given $c(t) \psi(0,t)$ should be identical to Eq.~\eqref{eq:formal2}.
At the insertion point $x=0$, it yields the following equation with
time-dependent coefficients:
\begin{equation}
\pi \psi_{tt}(0,t) + [10i\pi k^2 + 8ik c(t)] \psi_t (0,t) + [-9\pi k^4 -
40k^3 c(t) + 8ik c_t(t)] \psi(0,t) \approx 0.
\label{eq:two_mode}
\end{equation}
An advantage of this approximate formula is that $\psi(0,t)$ can be explicitly
obtained in terms of the Hermite polynomial and the hypergeometric function
for a simple protocol such as $c(t) \propto t$. However, the solution is not
very illuminating, and, more importantly, such a linear protocol does not split
the wave function within a finite time. So we will numerically integrate
Eq.~\eqref{eq:two_mode}, by choosing the protocol as $c(t) = \tan(k^2 t/4)$,
which diverges to infinity at $t^\ast \equiv 2\pi/k^2 = 8L^2/\pi$.
Starting from ground-state properties $\psi(0,0) = 1$ and $\psi_t(0,0) = -ik^2$
as the initial conditions, we see that Eq.~\eqref{eq:two_mode} now
describes amplitude changes as well (Fig.~\ref{fig:two_mode}). In particular,
the approximation predicts that the probability density $|\psi(0,t)|^2$
vanishes as $t$ approaches $t^\ast$, which is physically reasonable. It is also
plausible that the approximation can be improved systematically by including
more and more modes, and the number of included modes will correspond to the
order of the resulting ODE to integrate.
In the next section, we present a numerical method based on the
Riemann sum, by which one can carry out the summation of Eq.~\eqref{eq:withg}
over $\mu \lesssim O(10^2)$. Figure~\ref{fig:two_mode} shows that the
results with a few lowest modes are not very far apart from it.
We have so far discussed one way to use Eq.~\eqref{eq:two_mode}, i.e., starting
from $c(t)$ to obtain $\psi(0,t)$ and then proceed to $\psi(x,t)$. We may also
consider the opposite direction; that is, let us start from the desired
evolution of $\psi(x,t)$, from which $\psi(0,t)$ is obtained. If
Eq.~\eqref{eq:two_mode} is solved for $c(t)$ with this $\psi(0,t)$, the result is
the following (Appendix~\ref{appendix:ct}):
\begin{equation}
c(t) \approx \frac{iL e^{-5ik^2 t}}{4\psi(0,t)} \int_0^t dt' e^{5ik^2 t'}
\left( \frac{\partial}{\partial t'} +  ik^2 \right)
\left( \frac{\partial}{\partial t'} + 9ik^2 \right) \psi(0,t')
+ \frac{c(0) \psi(0,0)}{\psi(0,t)} e^{-5ik^2 t}.
\label{eq:ct}
\end{equation}
Equation~\eqref{eq:ct} suggests that one can design the protocol $c(t)$ to guide
the evolution of $\psi(x,t)$ within the two-mode approximation.
If $\psi(0,t)$ stays at one of the two lowest eigenmodes, for example, the first
term of Eq.~\eqref{eq:ct} identically vanishes and we find that $c(t) \approx
\frac{c(0) \psi(0,0)}{\psi(0,t)} e^{-5ik^2 t}$.
If $c(t)\neq 0$, one might reach a conclusion that $\psi(0,t)
\propto e^{-5ik^2 t}$, considering that $c(t)$ must be real for any $t$.
This is self-contradictory, however, because no eigenmode has phase velocity
$-5k^2$. The only possibility is to set $c(t)=0$. As a more nontrivial example,
let us write $\psi(0,t) = A(t) e^{i\varphi(t)}$ and suppose that the amplitude
decays as $A(t) = A_0 e^{-\lambda t}$ with real positive constants $A_0$ and
$\lambda$.
The phase $\varphi(t)$ is a real number in $(-\infty, +\infty)$.
Substituting this $\psi(0,t)$ into
Eq.~\eqref{eq:two_mode}, we obtain the following set of equations:
\begin{eqnarray}
&&\pi \varphi_t^2 + [10\pi k^2 + 8kc(t)] \varphi_t + 9\pi k^4 + 40k^3 c(t) - \pi
\lambda^2 = 0\label{eq:non1}\\
&&\pi \varphi_{tt} + 2\pi \lambda \varphi_t - 10\pi \lambda k^2 -8 \lambda kc(t)
+ 8kc_t(t) = 0\label{eq:non2}.
\end{eqnarray}
Equation~\eqref{eq:non1} is quadratic in $\varphi_t$ and can be solved as
\begin{equation}
\varphi_t = \frac{1}{2\pi} \left\{ -[10\pi k^2+8kc(t)] \pm \sqrt{64\pi^2 k^4 +
64k^2 c^2(t) + 4\pi^2 \lambda^2} \right\},
\label{eq:varphi}
\end{equation}
where we will choose the plus sign to have $\lim_{\lambda \rightarrow 0}
\varphi_t = -k^2$ on the ground state. By plugging Eq.~\eqref{eq:varphi} into
Eq.~\eqref{eq:non2}, one derives a nonlinear differential equation for $c(t)$.
It can be linearized by keeping only linear terms in $\lambda$ and $c(t)$,
which results in $c(t) \approx \frac{3\pi k}{4} (e^{4\lambda t}-1) \approx 3\pi
k \lambda t$. To sum up, if we choose how the amplitude changes over time,
it determines the evolution of the phase, due to the constraint that $c(t)$ must
real. By combining these $A(t)$ and $\varphi(t)$, it is possible to construct
$c(t)$. Note that the phase velocity $\varphi_t$ in Eq.~\eqref{eq:varphi}
converges to $-5k^2$ as $c(t) \rightarrow \infty$. This value turns out to be
an average of $-k^2$ and $-9k^2$, which are for the first and second modes,
respectively, and a little different from the phase velocity of the first
excited state, $-4k^2$, in this two-mode approximation.

\section{Numerical calculation}
\label{sec:numeric}

\begin{figure}
\includegraphics[width=0.45\textwidth]{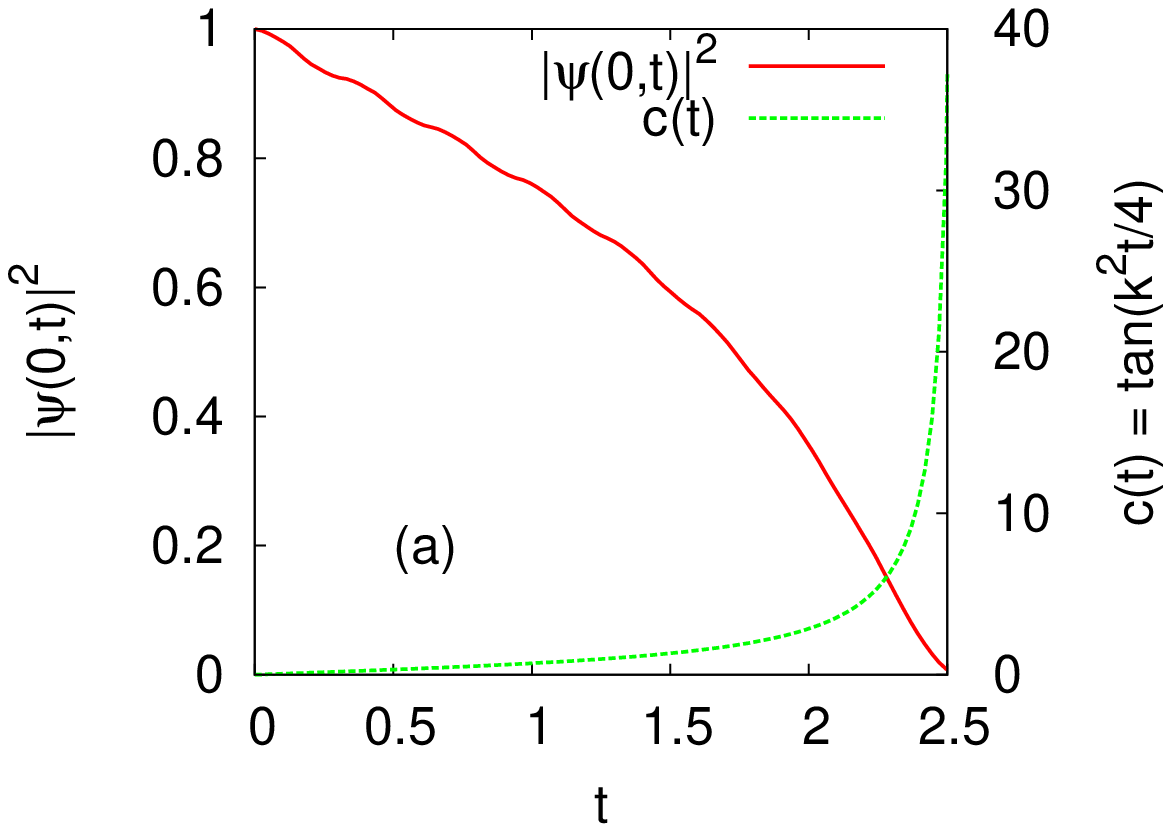}
\includegraphics[width=0.45\textwidth]{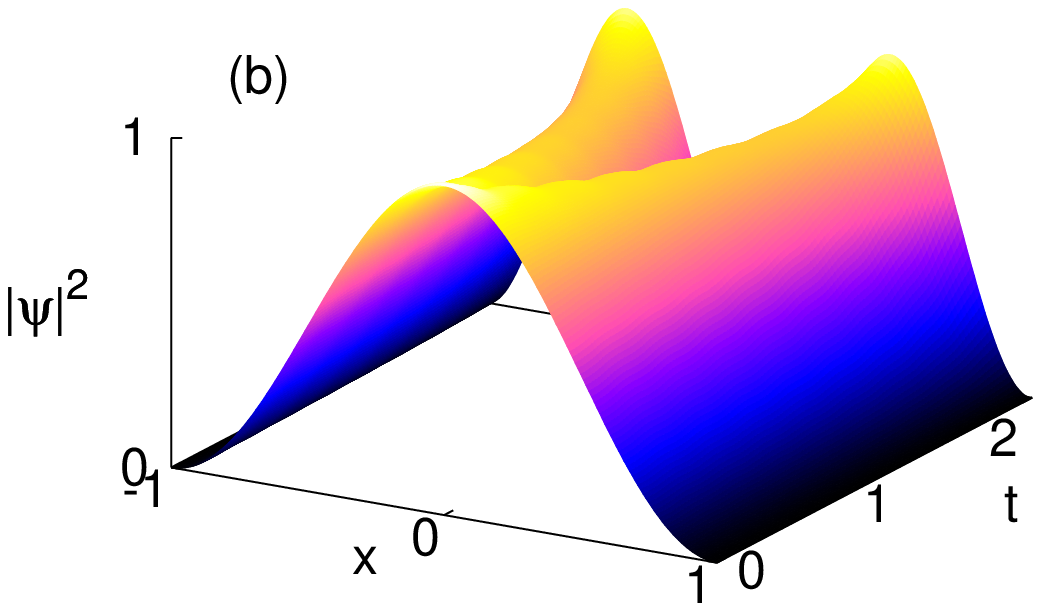}
\includegraphics[width=0.45\textwidth]{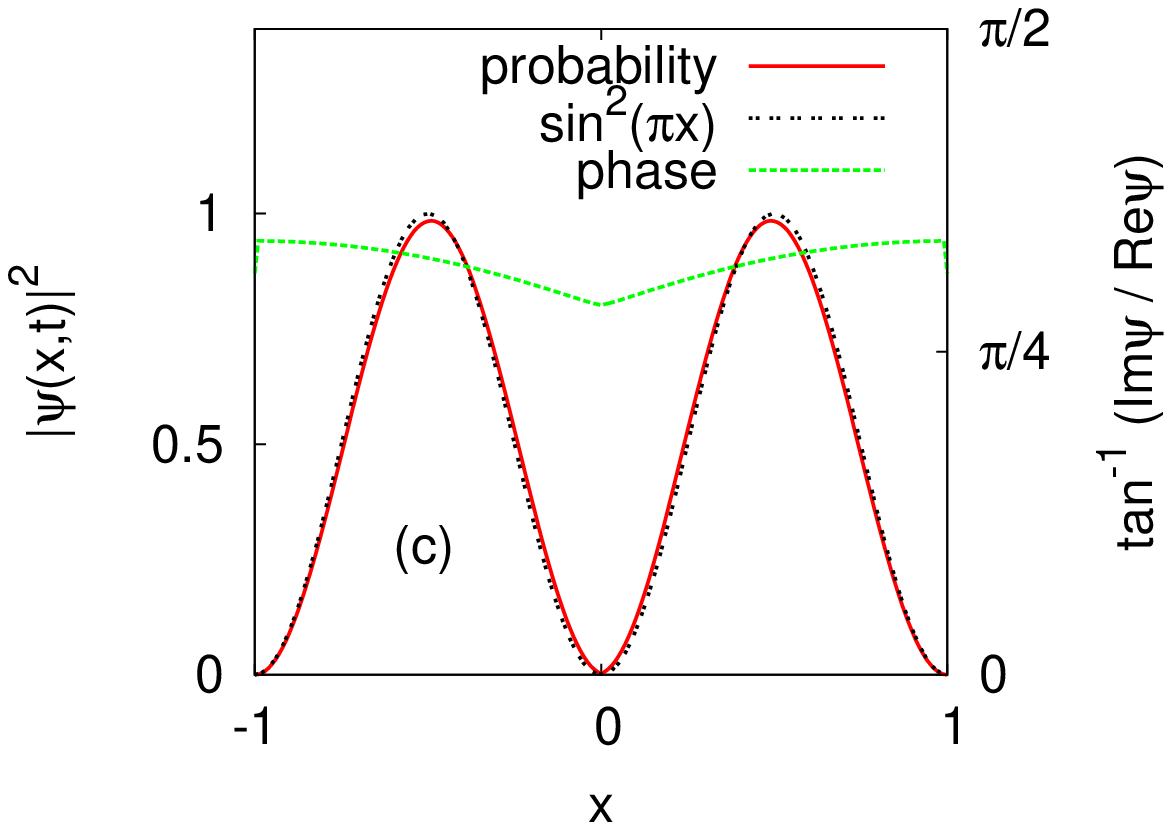}
\includegraphics[width=0.45\textwidth]{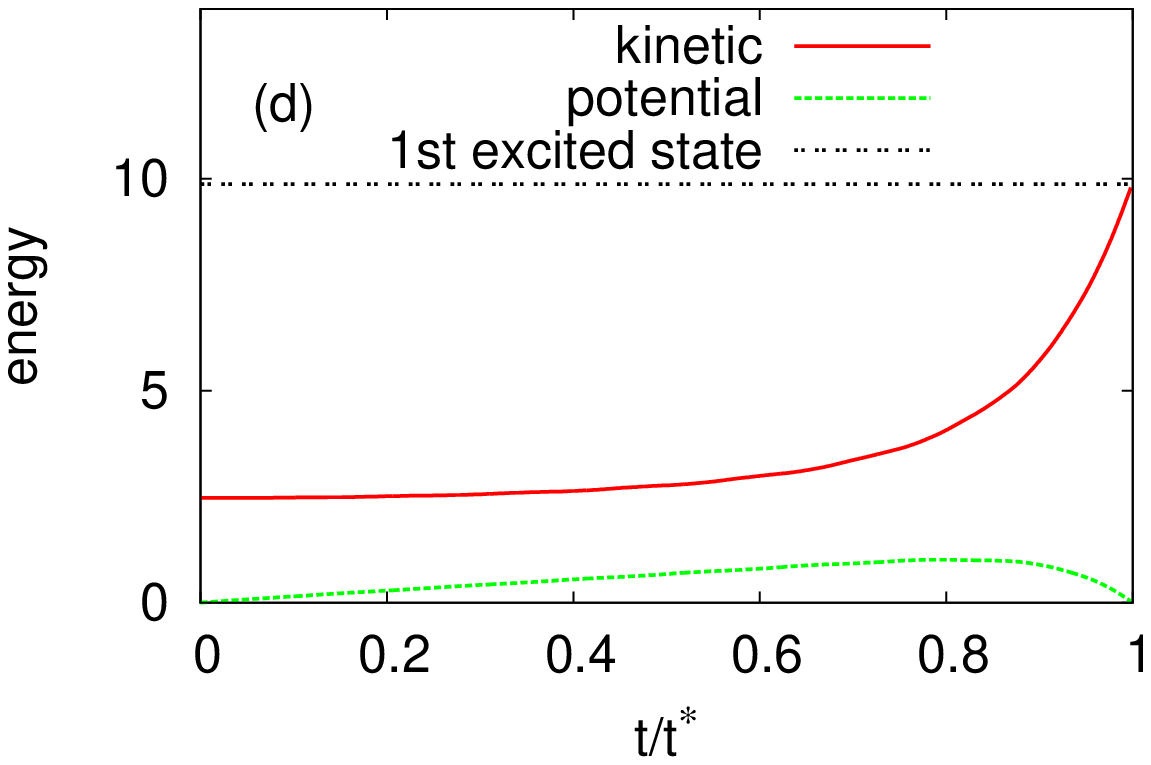}
\caption{(Color online) Numerical result of probability density $|\psi(x,t)|^2$
from an initial condition specified by Eq.~\eqref{eq:initial} with $L \equiv 1$.
The $\delta$ potential grows as $c(t) = \tan(k^2 t/4)$ at the origin and
diverges as $t \rightarrow t^\ast = 2\pi/k^2 = 8/\pi \approx 2.55$.
(a) Probability density at the origin as a function of
time, together with $c(t)$ for comparison.  (b) Full shape of
$|\psi(x,t)|^2$. (c) Probability density and phase at time $t =
0.99 \times t^\ast$ (solid). The black dotted line shows the probability density
of the first excited state, $\sin^2(\pi x)$, for comparison.
(d) Energy of the particle. The horizontal line indicates the energy of the
first excited state.
}
\label{fig:box}
\end{figure}

Numerical calculation can be a useful tool to explore
Eq.~\eqref{eq:volterra}, as we will show in this section.
Throughout this section, we will use the same protocol $c(t) = \tan(k^2 t/4)$
as in the above example. Recall that the box is completely separated into two
subsystems at $t = t^\ast = 2\pi/k^2$, so that we have to only consider $t \in
[0,t^\ast)$.
Our numerical strategy is to divide this time interval
into $M$ pieces to use the left Riemann sum for the
integral over $t'$ in Eq.~\eqref{eq:special}. It is convenient to define
\begin{equation}
F_\mu (t) \equiv \int_0^t dt' c(t') \psi(0,t') e^{i(2\mu+1)^2 k^2 t'},
\label{eq:fmu}
\end{equation}
and approximate it as
\begin{equation}
F_\mu (T\epsilon) \approx \epsilon \sum_{l=0}^{T-1} c(l\epsilon)
\psi(0,l\epsilon) e^{i(2\mu+1)^2 k^2 l\epsilon},
\end{equation}
where we have identified $dt$ and $t$ with $\epsilon$ and $T\epsilon$,
respectively, by introducing an integer $T \in (0,M)$
and $\epsilon \equiv t^\ast/M$.
At the next time step, $F_\mu$ is updated as follows:
\begin{equation}
F_\mu(T\epsilon+\epsilon) = F_\mu(T\epsilon) + \epsilon c(T\epsilon)
\psi(0,T\epsilon) e^{i(2\mu+1)^2 k^2 T\epsilon}.
\label{eq:fn}
\end{equation}
We obtain $\psi(0,T\epsilon)$ by calculating
\begin{equation}
\psi(0,T\epsilon) \approx L^{-1/2}e^{-ik^2 T\epsilon} - \frac{2i}{L}
\sum_{\mu=0}^{N} F_\mu(T\epsilon) e^{-i(2\mu+1)^2 k^2 T\epsilon}
\label{eq:psi0}
\end{equation}
with $N \gg 1$ to include a sufficiently large number of modes. At the same
time, it is worth noting that $(2N+1)^2 \epsilon$ must be small enough to
perform the integration in Eq.~\eqref{eq:fmu}.
By iterating Eqs.~\eqref{eq:fn} and \eqref{eq:psi0},
we can get $\psi(0,t)$ within the time interval $[0, t^\ast)$
[Fig.~\ref{fig:box}(a)].
Note that
Fig.~\ref{fig:two_mode} has already shown the curve of $|\psi(0,t)|^2$
obtained in this way to compare it with the two- and three-mode approximations.
We compute $\psi(x,T\epsilon)$ for nonzero $x$
in a similar manner, because Eq.~\eqref{eq:special} is discretized as
\begin{equation}
\psi(x,T\epsilon) \approx L^{-1/2}e^{-ik^2 T\epsilon} \cos kx - \frac{2i}{L}
\sum_{\mu=0}^{N} F_\mu(T\epsilon) e^{-i(2\mu+1)^2 k^2 T\epsilon}
\cos[(2\mu+1)kx].
\label{eq:numeric}
\end{equation}
Although we have discussed $x_0 = 0$ for brevity, the
generalization to $x_0 \neq 0$ is straightforward.
An advantage of this approach is that Eq.~\eqref{eq:numeric} can be computed
in parallel for many different $x$'s.
In this way, one can get the full shape of the probability density
$|\psi(x,t)|^2$ as depicted in Fig.~\ref{fig:box}(b).
Figure~\ref{fig:box}(c) shows $|\psi(x,t)|^2$ at $t = 0.99 \times t^\ast$ in
comparison with $\sin^2(\pi x/L)$, the probability density of the first excited
state. We have assured ourselves that the normalization condition
[Eq.~\eqref{eq:normalization}] is satisfied within the accuracy of $10^{-3}$
throughout the calculation.
Figure~\ref{fig:box}(d) shows the kinetic and potential energies of the particle
as functions of time [Eq.~\eqref{eq:energy}]. As $t$ approaches $t^\ast$, the
total energy converges to $4k^2$, the energy of the first excited state
$\phi_2(x)$.
As can be seen from Eq.~\eqref{eq:general}, however,
$\psi(x,t)$ must be symmetric with respect to the origin, so
we conclude that $\psi(x,t^\ast) \propto |\sin (2kx)|$ to a
good approximation.

Even if this is expected in the limit of the adiabatic
theorem~\cite{messiah}, it is not obvious {\it a priori}
when the barrier is inserted within finite time.
A qualitative explanation goes as follows:
The energy level spacing between the ground and
first excited states amounts to $3\pi^2/4 \approx 7.40$ at $t=0$ in our units.
The characteristic time scale will thus be of $O(4/3\pi^2) \sim O(10^{-1})$,
which is roughly $5\%$ of the total time $t^\ast \approx 2.55$ to insert the
barrier. For this reason, we can still say that the insertion is relatively
slow. This argument can be made more quantitative by approximating
our particle as a two-level system consisting of the ground state and
the first excited state.
Such approximation is reasonable, because other excited states are very far
away from these two in the energy spectrum.
According to the Landau-Zener formula, if the energy
levels approach to each other with speed $\dot\Delta$, the occupation
probability $P_2$ of the first excited state will be estimated by
$\ln(P_2) \propto - \dot\Delta^{-1}$ at the end of the protocol.
If $\dot\Delta \ll 1$, therefore, the
system will remain in the ground state with high probability, which is the case
when $t$ is roughly less than $0.8~t^\ast$ in Fig.~\ref{fig:box}(a).
The rapid increase of the barrier height for $t \gtrsim 0.8~t^\ast$
just gently pushes the particle out of the origin, because
the probability density at the origin has already become low at
$t = 0.8~t^\ast$. As a result, the particle is mostly preserved in the
ground state.

\begin{figure}
\includegraphics[width=0.45\textwidth]{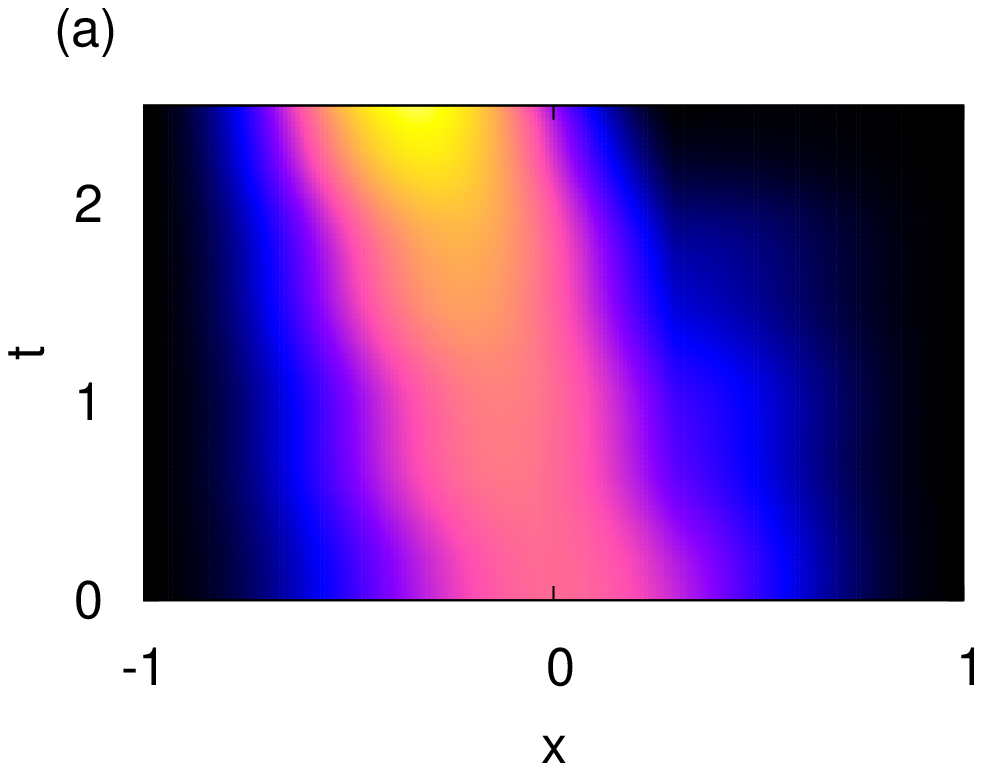}
\includegraphics[width=0.45\textwidth]{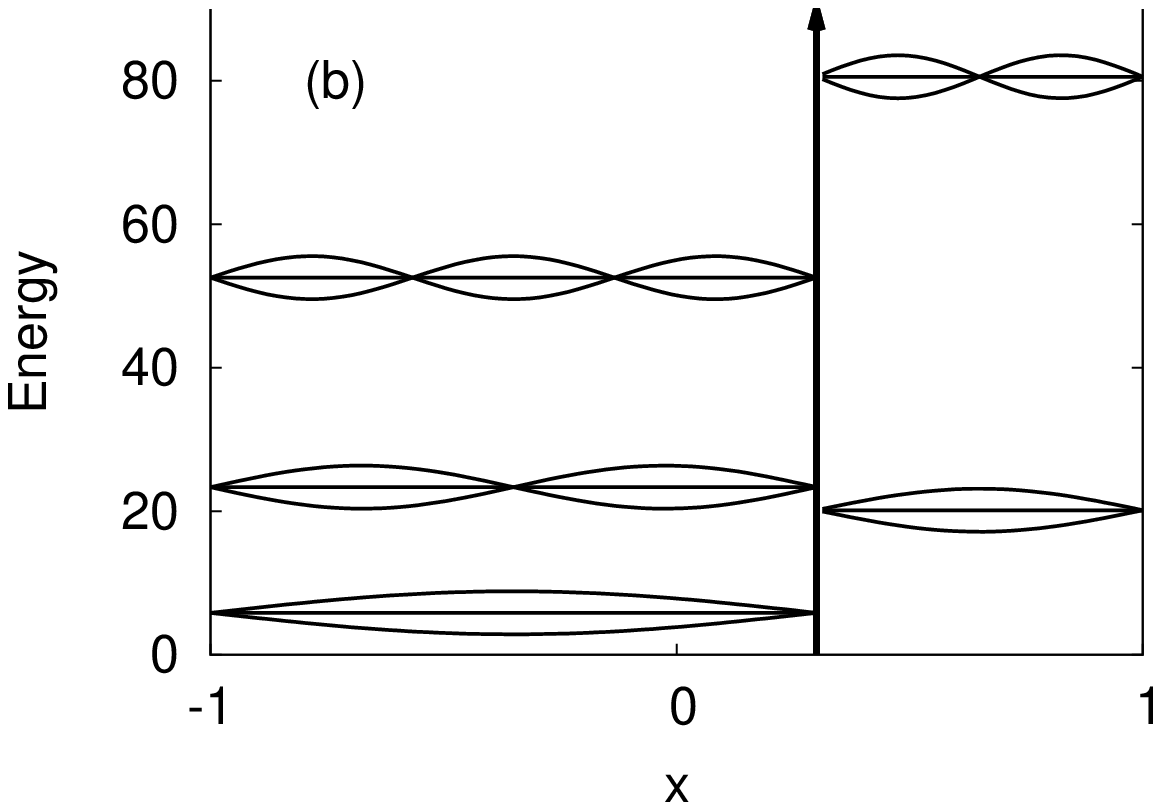}
\caption{(Color online) Time evolution of $|\psi(x,t)|^2$ when the barrier is
inserted at $x_0 = 0.3$.
(a) If the initial condition is the ground state, i.e., $\psi(x,0)=\phi_1(x) =
\cos kx$,
the particle will be localized on the left side of the box in the end.
(b) Energy levels in the two subsystems separated by the barrier at $x_0 =
0.3$, represented by the upward arrow.
The ground state of the total system is found on the left side.
}
\label{fig:asym}
\end{figure}

If the wall is away from the origin, i.e., $x_0 \neq 0$,
the particle can be localized at $t=t^\ast$ on the wider side of the
box~\cite{gea,*lakner,*enk,khkim}.
For example, Fig.~\ref{fig:asym}(a) shows the case of $x_0 = 0.3$.
This is also easily explained by applying the adiabatic theorem:
Let $L_1 = L + x_0$ denote the size of this wider side, whereas $L_2 = L-x_0$
denotes that of the other side.
The ground state at $t=t^\ast$ has
nonzero probability only on the wider side of the box with an
energy level $E_1 = \pi^2/L_1^2$ [Fig.~\ref{fig:asym}(b)].
The energy on the narrower side, $E_2 =
\pi^2/L_2^2$ is more than three times higher than this ground-state energy.
When the energy-level spacing is so large, our protocol can preserve the most
part of the system in the ground state.
We thus need only a single energy eigenstate to confine the
particle on one side of the box at $t=t^\ast$ if the barrier is slowly inserted
at an asymmetric position.
In other words, the which-side measurement commutes
with the Hamiltonian of this system in the quasistatic limit.
However, when the protocol is fast enough compared with the energy gap,
we can no longer neglect excitation, and this creates uncertainty
to be resolved by a which-side measurement.

\begin{figure}
\includegraphics[width=0.45\textwidth]{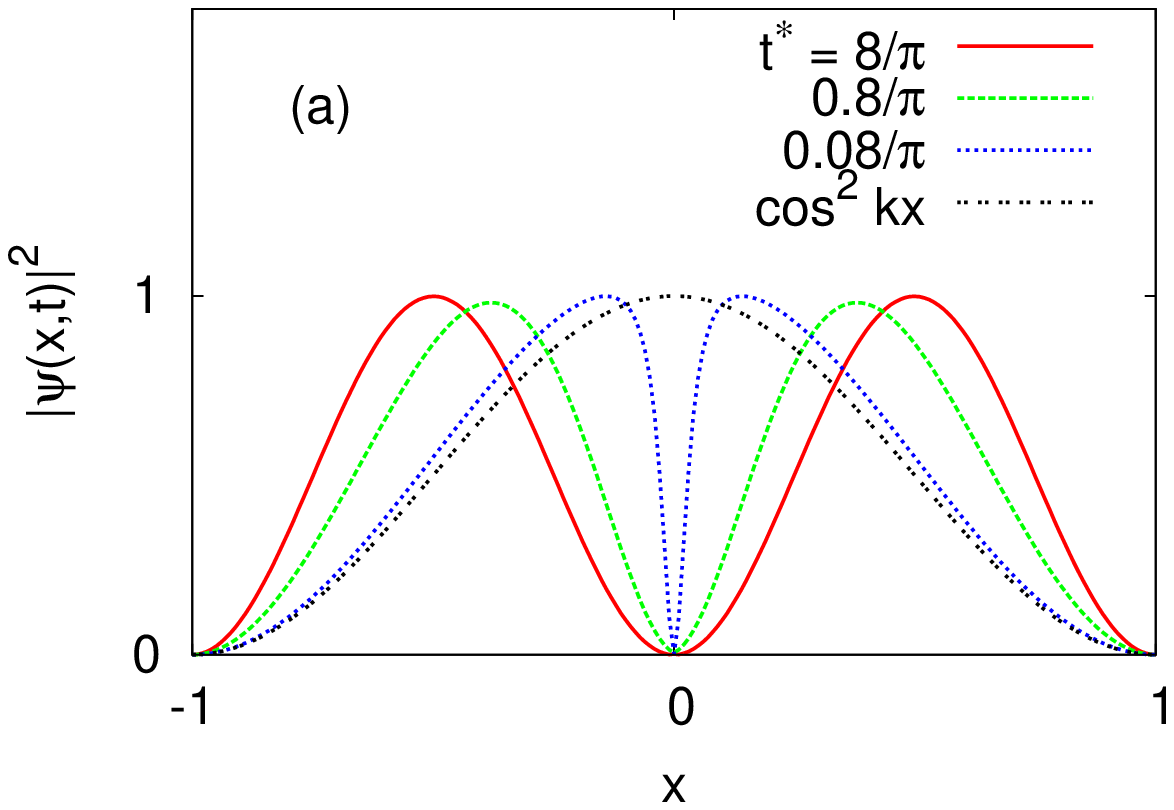}
\includegraphics[width=0.45\textwidth]{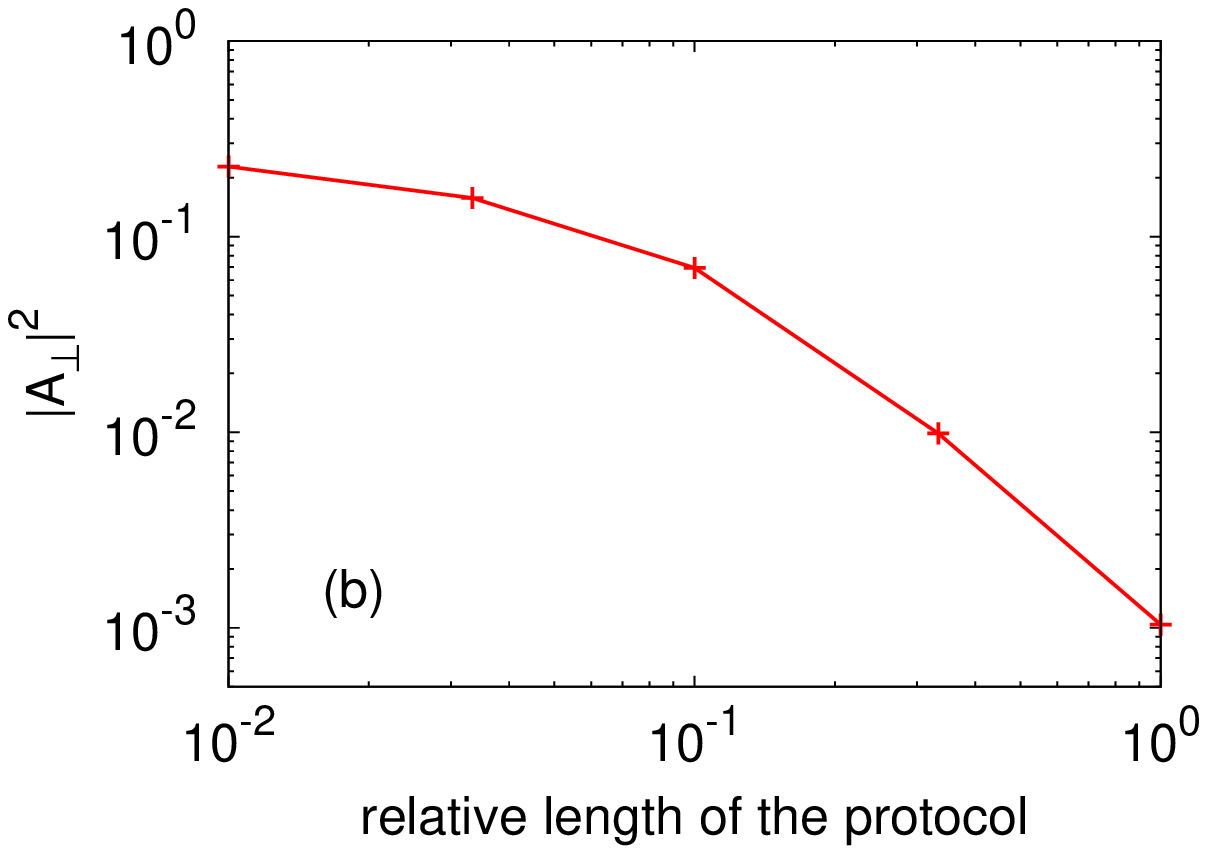}
\includegraphics[width=0.45\textwidth]{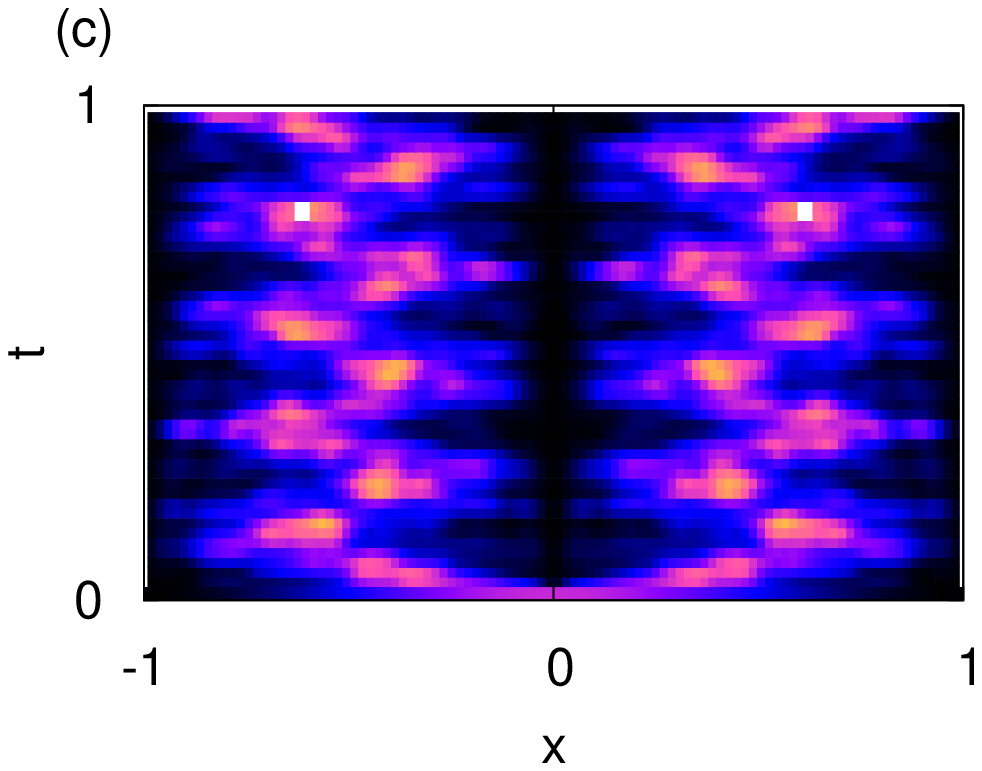}
\caption{(Color online) Insertion of the barrier at $x_0=0$ in a box with $L=1$.
The wave function is initially in the ground state, i.e., $\psi(x,0) = \phi_1
(x) = \cos kx$.
(a) Probability densities for various insertion speeds, when the wave function is
split into two parts. We have also plotted the initial probability density
function $|\psi(x,0)^2| = |\phi_1(x)|^2 = \cos^2 kx$ for comparison.
(b) Differences from the adiabatic limit (see text) near the end of the
protocol. On the horizontal axis, the unit length of the protocol is defined by
$8/\pi \approx 2.55$. (c) The time evolution of $|\psi(x,t)|^2$ in case of the
fastest insertion with $t^\ast=0.08/\pi$. The wave function for $t>t^\ast$
is obtained by using the spectral method~\cite{newman}. }
\label{fig:final}
\end{figure}

\begin{figure}
\includegraphics[width=0.45\textwidth]{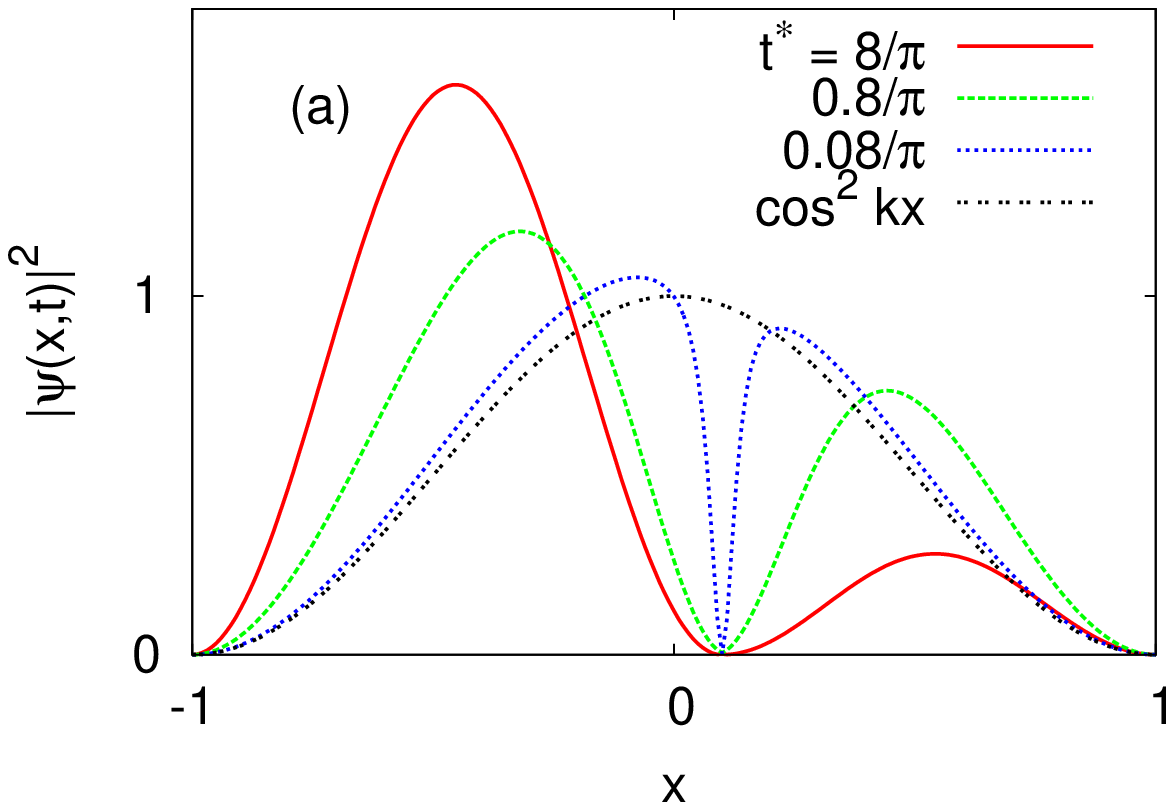}
\includegraphics[width=0.45\textwidth]{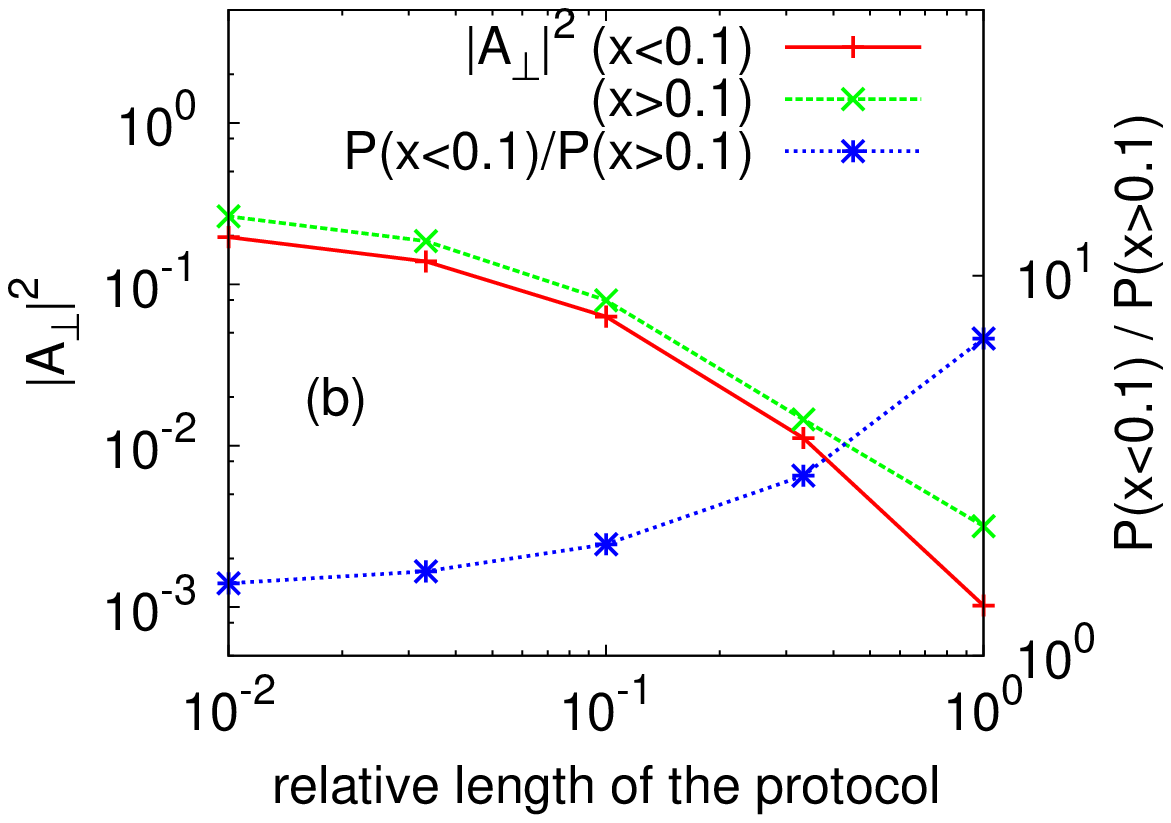}
\caption{(Color online) Evolution of the wave function for
various insertion speeds, when the barrier is inserted at $x_0 = 0.1$.
Except $x_0$, the other parameters are the same as in Fig.~\ref{fig:final}.
(a) Probability densities after the split, plotted with the initial probability
density function $|\psi(x,0)^2| = |\phi_1(x)|^2 = \cos^2 kx$ for comparison.
(b) Difference from the adiabatic limit on the two sides of the box at $t
\approx t^\ast$.
On the vertical secondary axis, we have also shown the ratio of
the probability to find the particle at $x<0.1$ to that of $x>0.1$.
}
\label{fig:uncertain}
\end{figure}

We have thus confirmed that our calculation produces physically reasonable
results as long as the process is slow enough with respect to the characteristic
time scale of the system. Obviously, our numerical calculation is not restricted
to such a slow protocol, and we can also check
what happens if we speed up the insertion. It is clear that the wave function
will not converge to the ground state of the subsystem any more but occupy
higher energy levels at the end of the process. Our calculation makes this guess
more precise: Figure~\ref{fig:final}(a) shows the probability densities at
$t \approx t^\ast$, and we have checked various $t^\ast$ across two orders of
magnitude.
The case of $t^\ast = 0.08/\pi$ describes a situation in which the insertion
speed is much faster in comparison to the characteristic time scale of the
system $\sim O(10^{-1})$.
As a consequence, the overall shape at the end of the protocol does
not deviate much from the starting point, $|\psi(x,0)|^2 = |\phi_1 (x)|^2 =
L^{-1} \cos^2 kx$.
As mentioned above, we expect $\psi(x,t) \rightarrow \psi_\parallel(x) \equiv
L^{-1/2} |\sin 2kx|$ in the adiabatic limit up to a phase factor.
We thus decompose $\psi(x,t)$ into parallel and perpendicular components to
$\psi_\parallel(x,t)$ as
\begin{equation}
\psi(x,t) = A_{\parallel} \psi_{\parallel}(x,t) + A_{\perp} \psi_{\perp}(x,t),
\end{equation}
where $\psi_\perp (x,t)$ is a normalized wave function perpendicular to
$\psi_\parallel(x,t)$. The normalization condition of $\psi(x,t)$ requires that
$|A_\parallel|^2 + |A_\perp|^2 = 1$. In Fig.~\ref{fig:final}(b), we plot
$|A_\perp|^2$ with varying the length of the protocol,
measured in units of $8/\pi$.
As we slow down the insertion, the difference from the
adiabatic limit decreases drastically. This point is especially important in
splitting a wave function within a finite
time~\cite{pezze,hohenester,*grond09a,*grond09b,masuda08,*masuda09,torrontegui}.
In case of a quick protocol, the excitation to higher-energy modes, as
manifested by the dip around the origin, makes the wave function rugged in
further time evolution [Fig.~\ref{fig:final}(c)]. We note that
this result is consistent with the predicted fractal
wave function~\cite{bender}.
Even if $x_0 \neq 0$, the excitation makes both sides of the box occupied,
creating uncertainty to be resolved by a which-side measurement
[Fig.~\ref{fig:uncertain}(a)]. Differently from the quasistatic protocol,
therefore, we can still make use of the information to extract work with a
finite-time protocol.
Our numerical procedure provides a precise way to obtain
the wave function, from which the amount of uncertainty can be estimated.
Figure~\ref{fig:uncertain}(b) depicts the trade-off between the uncertainty and
the excitation to other eigenmodes. This is crucial in the context of splitting
a matter wave, where it is desirable to keep the uncertainty while minimizing
excitation at the same time. Figure~\ref{fig:uncertain}(b) shows that the
excitation can be suppressed with $|A_\perp|^2 \lesssim O(10^{-2})$, while the
asymmetry of probabilities are also kept less than $O(10)$, if $t^\ast \approx
8/\pi$ for $x_0 = 0.1$.

\section{Discussion and Summary}
\label{sec:conclusion}

In summary, we have investigated a quantum particle in an isolated box with a
time-dependent $\delta$ potential.  We have found an integral expression for the
wave function [Eq.~\eqref{eq:volterra}] together with an approximate formula
[Eq.~\eqref{eq:two_mode}]. The numerical evaluation through
the integral equation gives precise results for the wave function during
the barrier insertion, even if the process is completed within finite time.
The total duration of the process, denoted by $t^\ast$, can be either short or
long compared to the characteristic time scale of the particle, and the
insertion point can be either symmetric or asymmetric.
For all the cases considered, our numerical calculation has produced
physically reasonable results in the light of the Landau-Zener formula.

In the context of splitting a wave function, we may consider a Bose-Einstein
condensate in an optical trap forming a one-dimensional
box~\cite{meyrath,henderson,zhang}. The $\delta$ potential barrier can be
implemented experimentally by shining a sharp laser beam onto the
condensate. If the barrier height is increased with a sufficiently slow rate
compared to the internal time scale of the system, the adiabatic limit can be
reproduced with good accuracy [Fig.~\ref{fig:final}(b)]. In addition,
we have also considered how to design the protocol within the two-mode
approximation [Eqs.~\eqref{eq:non1} and \eqref{eq:non2}].
If the barrier is off the middle of the box, its insertion localizes the
wave function on one side of the box, instead of splitting it, in the adiabatic
limit. This trade-off has been illustrated for $x_0 = 0.1$ in
Fig.~\ref{fig:uncertain}(b) so that one can choose the insertion speed
based on the calculation.

\begin{figure}
\includegraphics[width=0.45\textwidth]{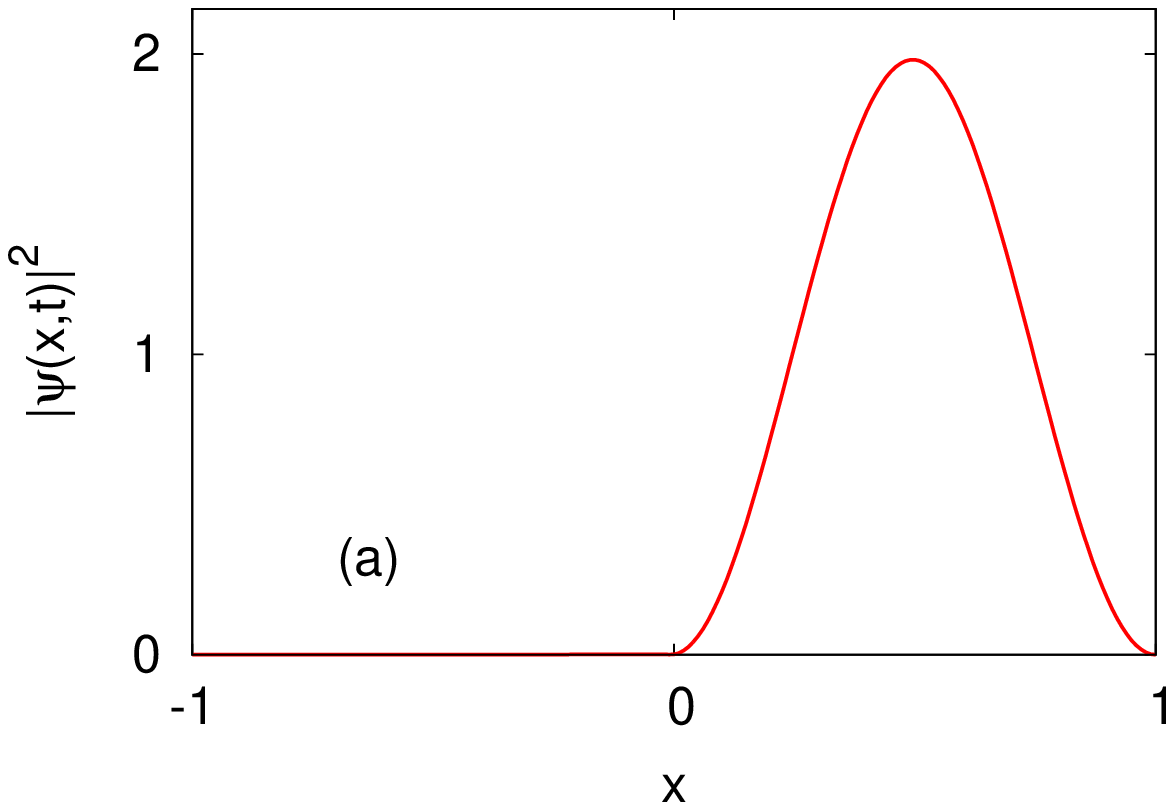}
\includegraphics[width=0.45\textwidth]{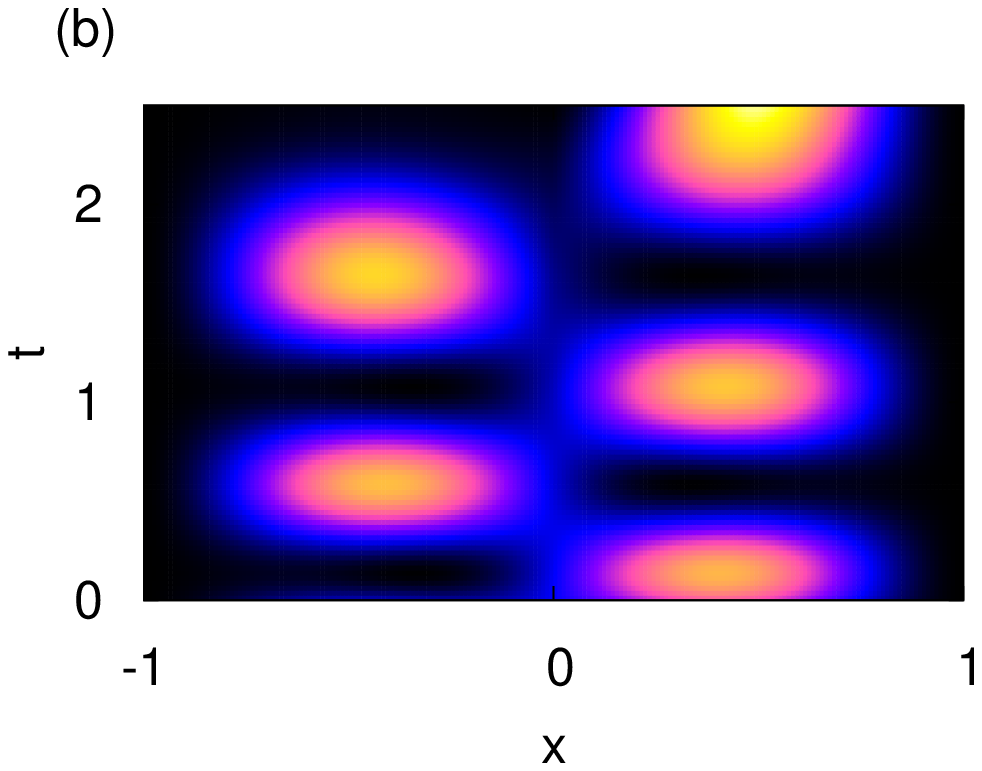}
\caption{(Color online) (a) Probability density of Eq.~\eqref{eq:Psi}.
at $t = 0.99 t^\ast$ when $t_0$ is suitably chosen. (b) Time evolution of
$|\Psi(x,t)|^2$ with the same $t_0$.}
\label{fig:tunnel}
\end{figure}

If we return to the Szilard engine, the insertion of the barrier is
followed by measuring the position of the particle and expanding this
single-particle gas isothermally. Thereafter, the barrier should be
removed to complete one cycle. As long as the engine can be treated as isolated
from the environment,
the removal of the barrier can be studied in the same way as in this work:
It boils down to the derivation of an integral equation similar to
Eq.~\eqref{eq:volterra}, beginning with a localized initial condition due to
the measurement.
Instead of starting from scratch, however, we point out that our results in the
previous section can be directly used to study the removal process in certain
cases. For example, suppose that the wave function is described in the form
\begin{equation}
\Psi(x,t) = \frac{1}{\sqrt{2}} \left[ \psi(x,t) + \phi_2(x,t) \right],
\label{eq:Psi}
\end{equation}
where $\psi(x,t)$ is obtained from Eq.~\eqref{eq:general} and
$\phi_2(x,t)$ is the first excited state including time dependence. We may write
it as $\phi_2(x,t) = \phi_2(x) e^{-4ik^2 (t-t_0)}$ with certain reference time
$t_0$.
It is obvious that $\phi_2(x,t)$ is not affected by the barrier,
because the barrier is located at a node of this wave function.
We also note that $\psi(x,t)$ and $\phi_2(x,t)$ are orthogonal to each other,
because one is even and the other is odd with respect to $x \rightarrow -x$.
By choosing a suitable $t_0$ for $\phi_2(x,t)$,
we can induce destructive interference on the left side of the box at
$t \approx t^\ast$, which effectively confines the particle to the right half of
the box [Fig.~\ref{fig:tunnel}(a)]. With this $t_0$, we can also obtain
the full time evolution of $\Psi(x,t)$ from $t=0$ to $t\approx t^\ast$
[Fig.~\ref{fig:tunnel}(b)].
If we define $\Psi^T(x,t) \equiv \overline{\Psi} \left(x,t^\ast-t \right)$,
where the overbar means complex conjugate, it satisfies the following equation:
\begin{equation}
-\Psi^{T}_{xx} + 2c(t^\ast-t)\delta(x-x_0) \Psi^{T} = i \Psi^T_t.
\end{equation}
Comparing this with the original Schr\"{o}dinger equation [Eq.~\eqref{eq:sch}],
we see that $\Psi^T(x,t)$ solves the case of time reversal.
Therefore, if we remove the barrier,
the probability density will evolve from the top to the bottom in
Fig.~\ref{fig:tunnel}(b), because $|\Psi^T(x,t)|^2 = |\Psi(x,t^\ast-t)|^2$.
We clearly observe tunneling of the particle,
resulting from a beat phenomenon between $\psi(x,t)$ and
$\phi_2(x,t)$~\cite{tunnel}, as the height of the barrier decreases.
Such a beat phenomenon is weakened when the barrier is off the origin, because
a single eigenstate is enough to localize a particle~[Fig.~\ref{fig:asym}(a)].

Considering that the quantum Szilard engine designed in Ref.~\onlinecite{swkim}
always keeps the quantum gas isothermal, it is fair to say that our
computational framework can only hint at low-temperature behavior at best,
because it deals with an isolated system. To understand the performance of
a quantum information engine working at a finite temperature, we need to take
into consideration a thermal reservoir in the quantum-mechanical context.
That is far beyond the scope of the present work and will be undertaken in a
future study.

\acknowledgments
We gratefully acknowledge discussions with Sang Wook Kim.
This work was supported by the Basic Science Research Program through the
National Research Foundation of Korea (NRF), funded by the Ministry of Science,
ICT and Future Planning (Grant No. NRF-2014R1A1A1003304).

\appendix

\section{Laplace transform of the Schr\"{o}dinger equation}
\label{appendix:laplace}

In the main text, we have written the Schr\"{o}dinger equation
through the Laplace transform as follows:
\begin{equation}
\overline\psi_{xx} + is \overline\psi = i\psi(x,0).
\end{equation}
The homogeneous equation in the absence of the RHS is solved
by
\begin{equation}
\overline\psi_c(x,s) = a(s) y_1(x) + b(s) y_2(x),
\label{eq:hom}
\end{equation}
where $y_1(x) \equiv e^{i\sqrt{is}x}$ and $y_2(x) \equiv e^{-i\sqrt{is}x}$.
The prefactors $a(s)$ and $b(s)$ are determined by the boundary condition.
The particular solution can be found by the integral
\begin{equation}
\overline\psi_p(x,s) = -y_1(x) \int dx' \frac{y_2(x')i\psi(x',0)}{W(x)} + y_2(x)
\int dx' \frac{y_1(x') i\psi(x',0)}{W(x)},
\end{equation}
where $W$ is the Wronskian of $y_1$ and $y_2$:
\begin{equation}
W =
\left| \begin{array}{cc}
y_1 & y_2 \\
y'_1 & y'_2
\end{array} \right|
=
\left| \begin{array}{cc}
e^{i\sqrt{is}x} & e^{-i\sqrt{is}x} \\
i\sqrt{is}e^{i\sqrt{is}x} & -i\sqrt{is} e^{-i\sqrt{is}x} \\
\end{array} \right|
= -2i \sqrt{is}.
\end{equation}
The solution of Eq.~\eqref{eq:lt} is given as
$\overline\psi(x,s) = \overline\psi_c(x,s) + \overline\psi_p(x,s)$.
More specifically, $\psi_p(x,s)$ can be expressed as
\begin{eqnarray}
\overline\psi_p(x,s) &=&
\int_{-\infty}^x dx' \frac{e^{i\sqrt{is}(x-x')}}{2\sqrt{is}} \psi(x',0) -
\int_{\infty}^x dx' \frac{e^{i\sqrt{is}(x'-x)}}{2\sqrt{is}} \psi(x',0)\\
&=&
\int_{-\infty}^{\infty} dx' \frac{e^{i\sqrt{is}|x-x'|}}{2\sqrt{is}} \psi(x',0).
\label{eq:particular}
\end{eqnarray}
For example, suppose that we choose the ground state as our initial condition:
\begin{equation}
\psi(x,0) = \left\{
\begin{array}{ll}
L^{-1/2} \cos kx & \mbox{if~}|x|<L,\\
0 & \mbox{otherwise,}
\end{array}
\right.
\label{eq:initial}
\end{equation}
where $k \equiv \pi/(2L)$.
The particular solution is then explicitly given as
\begin{equation}
\overline\psi_p(x,s) = \frac{
\sqrt{\frac{ik^2}{sL}} e^{i\sqrt{is}L} \cos \left(\sqrt{is}x \right) + L^{-1/2}
\cos kx} {s + ik^2}.
\end{equation}
We can carry out a similar integral to get an explicit expression
for $\overline\psi_p(x,s)$ by choosing our initial condition as the $n$th
eigenstate $\phi_n(x)$ in the absence of the $\delta$-function potential
\begin{equation}
\psi(x,0) = \phi_n(x) = \left\{
\begin{array}{ll}
L^{-1/2} \cos n k x & \mbox{if~}|x|<L \mbox{~and~$n$ is odd},\\
L^{-1/2} \sin n k x & \mbox{if~}|x|<L \mbox{~and~$n$ is even},\\
0 & \mbox{otherwise,}
\end{array}
\right.
\end{equation}
where $n=1$ corresponds to the ground state. To sum up,
Eq.~\eqref{eq:lt} has a solution of the following form:
\begin{equation}
\overline\psi(x,s) = \left\{
\begin{array}{ll}
a_+ e^{i\sqrt{is}x} + b_+ e^{-i\sqrt{is}x} + \overline\psi_p(x,s) & \mbox{if~}x>x_0,\\
a_- e^{i\sqrt{is}x} + b_- e^{-i\sqrt{is}x} + \overline\psi_p(x,s) & \mbox{if~}x<x_0.
\end{array}
\right.
\end{equation}
In addition, we have four conditions to determine the coefficients $a_\pm$ and
$b_\pm$:
\begin{eqnarray}
\lim_{x \rightarrow x_0^+} \overline\psi(x,s) = \lim_{x \rightarrow x_0^-}
\overline\psi(x,s),\\
\overline\psi(+L,s) = 0,\\
\overline\psi(-L,s) = 0,\\
\lim_{x \rightarrow x_0^+}\overline\psi_x (x,s) - \lim_{x \rightarrow
x_0^-}\overline\psi_x(x,s) = 2c(t) \psi(x_0,t).
\label{eq:barrier}
\end{eqnarray}
Note that the existence of the barrier inside the box is manifested by
the last boundary condition at $x=x_0$ [Eq.~\eqref{eq:barrier}]. One technical
remark is in order: When one solves the Schr\"{o}dinger equation with a
time-varying potential, it is also common to use the coupled-channel method,
which is perturbative expansion with the eigenmodes of the original barrier-free
system~\cite{gea}. However, the singular property of the $\delta$ function has
led us to treat the time-varying potential as a boundary condition, as expressed
in Eq.~\eqref{eq:barrier}. After some algebra, the solution is obtained as
\begin{equation}
\overline\psi(x,s) = \frac{\phi_n(x)}{s+in^2k^2} + \mathcal{L}[c(t) \psi(x_0,t)]
F(x,s),
\end{equation}
where
\begin{equation}
F(x,s) = \left\{
\begin{array}{ll}
\frac{\left( e^{2i\sqrt{is}L}-e^{2i\sqrt{is}x} \right)
\left( e^{2i\sqrt{is}(L+x_0)} - 1 \right)}
{i\sqrt{is} e^{i\sqrt{is}(x+x_0)} \left( e^{4i\sqrt{is}L} - 1 \right)}
& \mbox{for~} x_0 \le x < L,\\
\frac{\left( e^{2i\sqrt{is}L}-e^{2i\sqrt{is}x_0} \right)
\left( e^{2i\sqrt{is}(L+x)} - 1 \right)}
{i\sqrt{is} e^{i\sqrt{is}(x+x_0)} \left( e^{4i\sqrt{is}L} - 1 \right)}
& \mbox{for~} -L < x < x_0,
\end{array}
\right.
\end{equation}
which is the result in the main text.

If we choose a linear protocol such as $c(t) = c_0 t$ with a
real constant $c_0$, the
second term on the RHS of Eq.~\eqref{eq:psixt} has an explicit form because
$\mathcal{L}[t \times \psi(x_0,t)] = - \frac{d}{ds}
\overline{\psi}(x_0,s)$~\cite{campbell}.
For $x=x_0$, therefore, we get a first-order ODE,
\begin{equation}
\overline\psi(x_0,s) = \frac{\phi_n(x_0)}{s+in^2k^2} - c_0 \frac{d}{ds}
\overline{\psi}(x_0,s) F(x_0,s),
\end{equation}
whose formal solution is always available~\cite{boas}.
For example,
suppose that the particle occupies the ground state at $t=0$, i.e., $n=1$. If
the barrier is inserted at $x_0 = 0$, we find that
\begin{equation}
\frac{d}{ds} \overline{\psi}(0,s)
- c_0^{-1} \sqrt{is} \cot(\sqrt{is}L) \overline\psi(0,s)
= - c_0^{-1} \frac{L^{-1/2}\sqrt{is} \cot(\sqrt{is}L)}{s+in^2k^2}.
\end{equation}
The formal solution is thus written as
\begin{equation}
\overline\psi(0,s) = - c_0^{-1} e^{-I(s)} \int_0^s \frac{L^{-1/2}\sqrt{is'}
\cot(\sqrt{is'}L)}{s'+in^2k^2} e^{I(s')} ds' + \overline\psi(0,0) e^{-I(s)},
\end{equation}
where
\begin{eqnarray}
I(s) &\equiv& -c_0^{-1} \int_0^s \sqrt{is'} \cot(\sqrt{is'}L) ds'\nonumber\\
&=& \frac{2}{3} (is)^{3/2} + 2sL^{-1} \ln \left(1 - e^{-2i\sqrt{is}L}
\right) + 2L^{-2} \sqrt{is} \mbox{Li}_2 \left( e^{-2i\sqrt{is}L}
\right)\nonumber\\
&&
- iL^{-3} \mbox{Li}_3 \left( e^{-2i\sqrt{is}L} \right) + iL^{-3} \zeta(3).
\end{eqnarray}
Note that $\mbox{Li}_n$ is the polylogarithm function of order $n$,
and $\zeta$ is the Riemann $\zeta$ function. The next step would be to
use the solution $\overline\psi(x_0,s)$ to derive the full wave function
$\overline\psi(x,s)$.
Although the above procedure is formally exact, it is too complicated to obtain
the wave function in the $(x,t)$ space.
In addition, we are more interested in a general protocol, which may go up to
infinity within finite time. For this reason, we do not take this direction but
directly apply the inverse Laplace transform to Eq.~\eqref{eq:psixt} as shown
in the main text.

\section{Inverse Laplace transform of Eq.~\eqref{eq:psixt}}
\label{appendix:inverse}

Here, we consider the inverse Laplace transform of Eq.~\eqref{eq:psixt}.
It is elementary to transform the first term on the RHS of Eq.~\eqref{eq:psixt},
because $\mathcal{L} \left[ e^{-in^2 k^2 t} \right] = \frac{1}{s+in^2 k^2}$.
To use the convolution theorem for the second term, we need the
inverse Laplace transform of $F(x,s)$
through the Bromwich integral~\cite{boas}
\begin{equation}
\mathcal{L}^{-1} \left[ F(x,s) \right] =
\frac{1}{2\pi i} \int_{h-i\infty}^{h+i\infty} F(x,z) e^{zt} dz,
\label{eq:bromwich}
\end{equation}
where $t>0$.
\begin{figure}
\includegraphics[width=0.45\textwidth]{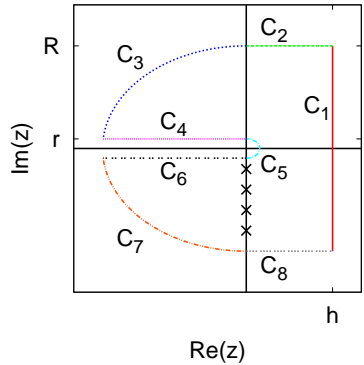}
\caption{(Color online) Keyhole contour to carry out the Bromwich integral
[Eq.~\eqref{eq:bromwich}]. The crosses on the imaginary axis represent the
poles of the integrand. The parameter $h$ is a positive number, which is kept
constant even when the radius $R$ of $C_3$ and $C_7$ grows to infinity. At the
same time, the radius of $C_5$, denoted as $r$, approaches zero.}
\label{fig:contour}
\end{figure}
Equation~\eqref{eq:Fs} has simple poles on the imaginary axis
at $z = -i \nu^2 k^2$ with $\nu=1, 2,\ldots$, which implies that
$h$ can be an arbitrary positive constant.
Let us consider a keyhole contour, as depicted in Fig.~\ref{fig:contour},
with taking the negative real axis as a branch cut.
The first path $C_1$ goes from $h-iR$ to
$h+iR$. If $R$ grows to infinity while $h$ is kept constant,
the integral along $C_1$ is directly related to Eq.~\eqref{eq:bromwich}.
In such a limiting process, the length of $C_2$ remains constant,
whereas the integrand decreases as $R^{-1/2}$. Therefore, we conclude that
the integral along $C_2$ vanishes as $R \rightarrow \infty$,
and the same conclusion holds for $C_8$.
Both $C_3$ and $C_7$ are ghost contours, because $\left| e^{zt}
\right| = \left| e^{Rt \cos\theta} \right|$ and $\cos \theta$ is negative for
$\theta \in (-\pi, \pi)$. One can also readily see that the integral along
$C_5$ can be made arbitrarily small by taking the radius $r \rightarrow 0$.
The remaining parts are $C_4$ and $C_6$.
Let us assume $x_0<x< L$, because there is little difference for
$-L<x<x_0$. Let $X$ denote $\operatorname{Re}(z)$.
We have $z = e^{i\pi}X = -X$ and $\sqrt{z} = e^{i\frac{\pi}{2}}
\sqrt{X} = i\sqrt{X}$ along $C_4$, whereas 
$z = e^{-i\pi}X = X$ and $\sqrt{z} = e^{-i\frac{\pi}{2}}
\sqrt{X} = -i\sqrt{X}$ along $C_6$.
Then, Eq.~\eqref{eq:Fs} gives us the integrands
\begin{equation}
\left. F(x,z) \right|_{C_4} =
\frac{\left( e^{-2\sqrt{iX}L}-e^{-2\sqrt{iX}x} \right)
\left( e^{-2\sqrt{iX}(L+x_0)} - 1 \right)}
{-\sqrt{iX} e^{-\sqrt{iX}(x+x_0)} \left( e^{-4\sqrt{iX}L} - 1 \right)},
\label{eq:c4}
\end{equation}
and
\begin{equation}
\left. F(x,z) \right|_{C_6} =
\frac{\left( e^{2\sqrt{iX}L}-e^{2\sqrt{iX}x} \right)
\left( e^{2\sqrt{iX}(L+x_0)} - 1 \right)}
{\sqrt{iX} e^{\sqrt{iX}(x+x_0)} \left( e^{4\sqrt{iX}L} - 1 \right)}.
\label{eq:c6}
\end{equation}
A little algebra shows that Eqs.~\eqref{eq:c4} and \eqref{eq:c6} are actually
identical. The integrals along $C_4$ and $C_6$ are in the opposite directions,
and therefore cancel out each other. Now we have evaluated the integral around
the whole keyhole contour, and the residue theorem gives us $f_\nu(x,t)$ for
each pole as follows:
\begin{equation}
f_\nu (x,t) = \left\{
\begin{array}{ll}
\frac{1}{2iL} 
e^{-i\nu^2 k^2 t - i\nu k(x+x_0)} \left[ (-1)^\nu - e^{2i\nu kx} \right] \left[
e^{2i \nu k(L+x_0)} -1 \right] & \mbox{for~} x_0 < x< L\\
\frac{1}{2iL} 
e^{-i\nu^2 k^2 t - i\nu k(x+x_0)} \left[ (-1)^\nu - e^{2i\nu kx_0} \right]\left[
e^{2i \nu k(L+x)} -1 \right] & \mbox{for~} -L < x< x_0.
\end{array} \right.
\end{equation}
One might therefore conclude that $\mathcal{L}^{-1}[F(x,s)] =
\sum_{\nu=1}^\infty f_\nu(x,t)$, but the summation should be understood as a
formal expansion, whose convergence is not guaranteed. Based on the existence of
the wave function, we conjecture that the summation over $\nu$ will be generally
convergent after integrated over $t$ in Eq.~\eqref{eq:volterra}.
Unfortunately, it is difficult to prove this statement, because the integral of
Eq.~\eqref{eq:volterra} is also involved in $\psi(x_0,t)$, which is unknown as
yet. Still, we suggest that $\sin \nu t$ could be an illustrative example: It
does not converge when summed over $\nu=1,2,\ldots$. However, if we first
integrate it over $t$ and then carry out the summation, the result can converge
to a well-defined value.

\section{Solving Eq.~\eqref{eq:two_mode} for $c(t)$}
\label{appendix:ct}

Let us rearrange the terms of Eq.~\eqref{eq:two_mode} as follows:
\begin{equation}
8ik \psi(0,t) c_t(t) + \left[ 8ik \psi_t (0,t) - 40k^3 \psi(0,t) \right] c(t) +
\pi \left[ \psi_{tt}(0,t) + 10i k^2 \psi_t (0,t) - 9k^4 \psi(0,t) \right]
\approx 0.
\end{equation}
Dividing both sides by $8ik\psi(0,t)$, we get
\begin{equation}
c_t(t) + \left[ \frac{\psi_t (0,t)}{\psi(0,t)} + 5ik^2 \right] c(t) \approx
\frac{iL}{4\psi(0,t)} \left[ \psi_{tt}(0,t) + 10i k^2 \psi_t (0,t) - 9k^4
\psi(0,t) \right],
\end{equation}
which is a linear first-order ODE for $c(t)$. Its formal solution takes the
form~\cite{boas}
\begin{equation}
c(t) = e^{-I(t)} \int_0^t dt' Q(t') e^{I(t')} + c(0) e^{-I(t)},
\label{eq:formal1}
\end{equation}
where
\begin{eqnarray}
I(t) &\equiv& \int_0^t dt' \left[ \frac{\psi_t (0,t)}{\psi(0,t)} + 5ik^2
\right]\nonumber\\
&=& \int_0^t dt' \left\{ \frac{d}{dt'} \left[ \ln \psi(0,t')
\right] + 5ik^2 \right\}\nonumber\\
&=& \ln \psi(0,t) - \ln \psi(0,0) + 5ik^2 t,
\label{eq:it}
\end{eqnarray}
and $Q(t) \equiv
\frac{iL}{4\psi(0,t)} \left[ \psi_{tt}(0,t) + 10i k^2 \psi_t (0,t) - 9k^4
\psi(0,t) \right]$. If we plug Eq.~\eqref{eq:it} into Eq.~\eqref{eq:formal1},
the first term on the RHS of Eq.~\eqref{eq:formal1} is evaluated as
\begin{eqnarray}
e^{-I(t)} \int_0^t dt' Q(t') e^{I(t')}
&=& \frac{iL e^{-5ik^2t}}{4\psi(0,t)}
\int_0^t dt' e^{5ik^2 t'} \left( \frac{\partial}{\partial t'} + ik^2 \right)
\left( \frac{\partial}{\partial t'} + 9ik^2 \right) \psi(0,t'),
\end{eqnarray}
and the second term gives
\begin{equation}
c(0) e^{-I(t)} = \frac{c(0) \psi(0,0)}{\psi(0,t)} e^{-5ik^2 t}.
\end{equation}


\begin{thebibliography}{49}%
\makeatletter
\providecommand \@ifxundefined [1]{%
 \@ifx{#1\undefined}
}%
\providecommand \@ifnum [1]{%
 \ifnum #1\expandafter \@firstoftwo
 \else \expandafter \@secondoftwo
 \fi
}%
\providecommand \@ifx [1]{%
 \ifx #1\expandafter \@firstoftwo
 \else \expandafter \@secondoftwo
 \fi
}%
\providecommand \natexlab [1]{#1}%
\providecommand \enquote  [1]{``#1''}%
\providecommand \bibnamefont  [1]{#1}%
\providecommand \bibfnamefont [1]{#1}%
\providecommand \citenamefont [1]{#1}%
\providecommand \href@noop [0]{\@secondoftwo}%
\providecommand \href [0]{\begingroup \@sanitize@url \@href}%
\providecommand \@href[1]{\@@startlink{#1}\@@href}%
\providecommand \@@href[1]{\endgroup#1\@@endlink}%
\providecommand \@sanitize@url [0]{\catcode `\\12\catcode `\$12\catcode
  `\&12\catcode `\#12\catcode `\^12\catcode `\_12\catcode `\%12\relax}%
\providecommand \@@startlink[1]{}%
\providecommand \@@endlink[0]{}%
\providecommand \url  [0]{\begingroup\@sanitize@url \@url }%
\providecommand \@url [1]{\endgroup\@href {#1}{\urlprefix }}%
\providecommand \urlprefix  [0]{URL }%
\providecommand \Eprint [0]{\href }%
\providecommand \doibase [0]{http://dx.doi.org/}%
\providecommand \selectlanguage [0]{\@gobble}%
\providecommand \bibinfo  [0]{\@secondoftwo}%
\providecommand \bibfield  [0]{\@secondoftwo}%
\providecommand \translation [1]{[#1]}%
\providecommand \BibitemOpen [0]{}%
\providecommand \bibitemStop [0]{}%
\providecommand \bibitemNoStop [0]{.\EOS\space}%
\providecommand \EOS [0]{\spacefactor3000\relax}%
\providecommand \BibitemShut  [1]{\csname bibitem#1\endcsname}%
\let\auto@bib@innerbib\@empty
\bibitem [{\citenamefont {Alivisatos}(1996)}]{alivisatos}%
  \BibitemOpen
  \bibfield  {author} {\bibinfo {author} {\bibfnamefont {A.~P.}\ \bibnamefont
  {Alivisatos}},\ }\href@noop {} {\bibfield  {journal} {\bibinfo  {journal}
  {Science}\ }\textbf {\bibinfo {volume} {271}},\ \bibinfo {pages} {933}
  (\bibinfo {year} {1996})}\BibitemShut {NoStop}%
\bibitem [{\citenamefont {Swendsen}(2012)}]{swendsen}%
  \BibitemOpen
  \bibfield  {author} {\bibinfo {author} {\bibfnamefont {R.~H.}\ \bibnamefont
  {Swendsen}},\ }\href@noop {} {\emph {\bibinfo {title} {An introduction to
  Statistical Mechanics and Thermodynamics}}}\ (\bibinfo  {publisher} {Oxford
  University Press},\ \bibinfo {address} {Oxford},\ \bibinfo {year}
  {2012})\BibitemShut {NoStop}%
\bibitem [{\citenamefont {Kittel}(2005)}]{kittel}%
  \BibitemOpen
  \bibfield  {author} {\bibinfo {author} {\bibfnamefont {C.}~\bibnamefont
  {Kittel}},\ }\href@noop {} {\emph {\bibinfo {title} {Introduction to Solid
  State Physics}}}\ (\bibinfo  {publisher} {John Wiley \& Sons},\ \bibinfo
  {address} {Hoboken, NJ},\ \bibinfo {year} {2005})\BibitemShut {NoStop}%
\bibitem [{\citenamefont {Szilard}(1929)}]{szilard}%
  \BibitemOpen
  \bibfield  {author} {\bibinfo {author} {\bibfnamefont {L.}~\bibnamefont
  {Szilard}},\ }\href@noop {} {\bibfield  {journal} {\bibinfo  {journal} {Z.
  Phys.}\ }\textbf {\bibinfo {volume} {53}},\ \bibinfo {pages} {840} (\bibinfo
  {year} {1929})}\BibitemShut {NoStop}%
\bibitem [{\citenamefont {Zurek}(1986)}]{zurek}%
  \BibitemOpen
  \bibfield  {author} {\bibinfo {author} {\bibfnamefont {W.}~\bibnamefont
  {Zurek}},\ }in\ \href@noop {} {\emph {\bibinfo {booktitle} {Frontiers of
  Nonequilibrium Statistical Physics}}},\ \bibinfo {series} {NATO ASI Series},
  Vol.\ \bibinfo {volume} {135},\ \bibinfo {editor} {edited by\ \bibinfo
  {editor} {\bibfnamefont {G.~T.}\ \bibnamefont {Moore}}\ and\ \bibinfo
  {editor} {\bibfnamefont {M.~O.}\ \bibnamefont {Scully}}}\ (\bibinfo
  {publisher} {Plenum},\ \bibinfo {address} {New York},\ \bibinfo {year}
  {1986})\ pp.\ \bibinfo {pages} {151--161}\BibitemShut {NoStop}%
\bibitem [{\citenamefont {Kim}\ \emph {et~al.}(2011)\citenamefont {Kim},
  \citenamefont {Sagawa}, \citenamefont {{De Liberato}},\ and\ \citenamefont
  {Ueda}}]{swkim}%
  \BibitemOpen
  \bibfield  {author} {\bibinfo {author} {\bibfnamefont {S.~W.}\ \bibnamefont
  {Kim}}, \bibinfo {author} {\bibfnamefont {T.}~\bibnamefont {Sagawa}},
  \bibinfo {author} {\bibfnamefont {S.}~\bibnamefont {{De Liberato}}}, \ and\
  \bibinfo {author} {\bibfnamefont {M.}~\bibnamefont {Ueda}},\ }\href@noop {}
  {\bibfield  {journal} {\bibinfo  {journal} {Phys. Rev. Lett.}\ }\textbf
  {\bibinfo {volume} {106}},\ \bibinfo {pages} {070401} (\bibinfo {year}
  {2011})}\BibitemShut {NoStop}%
\bibitem [{\citenamefont {Sagawa}\ and\ \citenamefont {Ueda}(2008)}]{sagawa08}%
  \BibitemOpen
  \bibfield  {author} {\bibinfo {author} {\bibfnamefont {T.}~\bibnamefont
  {Sagawa}}\ and\ \bibinfo {author} {\bibfnamefont {M.}~\bibnamefont {Ueda}},\
  }\href@noop {} {\bibfield  {journal} {\bibinfo  {journal} {Phys. Rev. Lett.}\
  }\textbf {\bibinfo {volume} {100}},\ \bibinfo {pages} {080403} (\bibinfo
  {year} {2008})}\BibitemShut {NoStop}%
\bibitem [{\citenamefont {Jacobs}(2009)}]{jacobs2}%
  \BibitemOpen
  \bibfield  {author} {\bibinfo {author} {\bibfnamefont {K.}~\bibnamefont
  {Jacobs}},\ }\href@noop {} {\bibfield  {journal} {\bibinfo  {journal} {Phys.
  Rev. A}\ }\textbf {\bibinfo {volume} {80}},\ \bibinfo {pages} {012322}
  (\bibinfo {year} {2009})}\BibitemShut {NoStop}%
\bibitem [{\citenamefont {Toyabe}\ \emph {et~al.}(2010)\citenamefont {Toyabe},
  \citenamefont {Sagawa}, \citenamefont {Ueda}, \citenamefont {Muneyuki},\ and\
  \citenamefont {Sano}}]{toyabe}%
  \BibitemOpen
  \bibfield  {author} {\bibinfo {author} {\bibfnamefont {S.}~\bibnamefont
  {Toyabe}}, \bibinfo {author} {\bibfnamefont {T.}~\bibnamefont {Sagawa}},
  \bibinfo {author} {\bibfnamefont {M.}~\bibnamefont {Ueda}}, \bibinfo {author}
  {\bibfnamefont {E.}~\bibnamefont {Muneyuki}}, \ and\ \bibinfo {author}
  {\bibfnamefont {M.}~\bibnamefont {Sano}},\ }\href@noop {} {\bibfield
  {journal} {\bibinfo  {journal} {Nat. Phys.}\ }\textbf {\bibinfo {volume}
  {6}},\ \bibinfo {pages} {988} (\bibinfo {year} {2010})}\BibitemShut {NoStop}%
\bibitem [{\citenamefont {Sagawa}\ and\ \citenamefont {Ueda}(2010)}]{sagawa10}%
  \BibitemOpen
  \bibfield  {author} {\bibinfo {author} {\bibfnamefont {T.}~\bibnamefont
  {Sagawa}}\ and\ \bibinfo {author} {\bibfnamefont {M.}~\bibnamefont {Ueda}},\
  }\href@noop {} {\bibfield  {journal} {\bibinfo  {journal} {Phys. Rev. Lett.}\
  }\textbf {\bibinfo {volume} {104}},\ \bibinfo {pages} {090602} (\bibinfo
  {year} {2010})}\BibitemShut {NoStop}%
\bibitem [{\citenamefont {Sagawa}\ and\ \citenamefont
  {Ueda}(2012{\natexlab{a}})}]{sagawa12}%
  \BibitemOpen
  \bibfield  {author} {\bibinfo {author} {\bibfnamefont {T.}~\bibnamefont
  {Sagawa}}\ and\ \bibinfo {author} {\bibfnamefont {M.}~\bibnamefont {Ueda}},\
  }\href@noop {} {\bibfield  {journal} {\bibinfo  {journal} {Phys. Rev. Lett.}\
  }\textbf {\bibinfo {volume} {109}},\ \bibinfo {pages} {180602} (\bibinfo
  {year} {2012}{\natexlab{a}})}\BibitemShut {NoStop}%
\bibitem [{\citenamefont {Sagawa}\ and\ \citenamefont
  {Ueda}(2012{\natexlab{b}})}]{sagawa12b}%
  \BibitemOpen
  \bibfield  {author} {\bibinfo {author} {\bibfnamefont {T.}~\bibnamefont
  {Sagawa}}\ and\ \bibinfo {author} {\bibfnamefont {M.}~\bibnamefont {Ueda}},\
  }\href@noop {} {\bibfield  {journal} {\bibinfo  {journal} {Phys. Rev. E}\
  }\textbf {\bibinfo {volume} {85}},\ \bibinfo {pages} {021104} (\bibinfo
  {year} {2012}{\natexlab{b}})}\BibitemShut {NoStop}%
\bibitem [{\citenamefont {Sagawa}\ and\ \citenamefont {Ueda}(2013)}]{sagawa13}%
  \BibitemOpen
  \bibfield  {author} {\bibinfo {author} {\bibfnamefont {T.}~\bibnamefont
  {Sagawa}}\ and\ \bibinfo {author} {\bibfnamefont {M.}~\bibnamefont {Ueda}},\
  }in\ \href@noop {} {\emph {\bibinfo {booktitle} {Nonequilibrium Statistical
  Physics of Small Systems: Fluctuation Relations and Beyond}}},\ \bibinfo
  {editor} {edited by\ \bibinfo {editor} {\bibfnamefont {R.}~\bibnamefont
  {Klages}}, \bibinfo {editor} {\bibfnamefont {W.}~\bibnamefont {Just}}, \ and\
  \bibinfo {editor} {\bibfnamefont {C.}~\bibnamefont {Jarzynski}}}\ (\bibinfo
  {publisher} {Wiley},\ \bibinfo {address} {Weinheim},\ \bibinfo {year}
  {2013})\ pp.\ \bibinfo {pages} {181--211}\BibitemShut {NoStop}%
\bibitem [{\citenamefont {Jacobs}(2012)}]{jacobs1}%
  \BibitemOpen
  \bibfield  {author} {\bibinfo {author} {\bibfnamefont {K.}~\bibnamefont
  {Jacobs}},\ }\href@noop {} {\bibfield  {journal} {\bibinfo  {journal} {Phys.
  Rev. E}\ }\textbf {\bibinfo {volume} {86}},\ \bibinfo {pages} {040106(R)}
  (\bibinfo {year} {2012})}\BibitemShut {NoStop}%
\bibitem [{\citenamefont {Horowitz}\ \emph {et~al.}(2013)\citenamefont
  {Horowitz}, \citenamefont {Sagawa},\ and\ \citenamefont
  {Parrondo}}]{horowitz}%
  \BibitemOpen
  \bibfield  {author} {\bibinfo {author} {\bibfnamefont {J.~M.}\ \bibnamefont
  {Horowitz}}, \bibinfo {author} {\bibfnamefont {T.}~\bibnamefont {Sagawa}}, \
  and\ \bibinfo {author} {\bibfnamefont {J.~M.~R.}\ \bibnamefont {Parrondo}},\
  }\href@noop {} {\bibfield  {journal} {\bibinfo  {journal} {Phys. Rev. Lett.}\
  }\textbf {\bibinfo {volume} {111}},\ \bibinfo {pages} {010602} (\bibinfo
  {year} {2013})}\BibitemShut {NoStop}%
\bibitem [{\citenamefont {Um}\ \emph {et~al.}(2015)\citenamefont {Um},
  \citenamefont {Hinrichsen}, \citenamefont {Kwon},\ and\ \citenamefont
  {Park}}]{um}%
  \BibitemOpen
  \bibfield  {author} {\bibinfo {author} {\bibfnamefont {J.}~\bibnamefont
  {Um}}, \bibinfo {author} {\bibfnamefont {H.}~\bibnamefont {Hinrichsen}},
  \bibinfo {author} {\bibfnamefont {C.}~\bibnamefont {Kwon}}, \ and\ \bibinfo
  {author} {\bibfnamefont {H.}~\bibnamefont {Park}},\ }\href@noop {} {\bibfield
   {journal} {\bibinfo  {journal} {New J. Phys.}\ }\textbf {\bibinfo {volume}
  {17}},\ \bibinfo {pages} {085001} (\bibinfo {year} {2015})}\BibitemShut
  {NoStop}%
\bibitem [{\citenamefont {Park}\ \emph {et~al.}(2015)\citenamefont {Park},
  \citenamefont {Yi},\ and\ \citenamefont {Baek}}]{park}%
  \BibitemOpen
  \bibfield  {author} {\bibinfo {author} {\bibfnamefont {M.}~\bibnamefont
  {Park}}, \bibinfo {author} {\bibfnamefont {S.~D.}\ \bibnamefont {Yi}}, \ and\
  \bibinfo {author} {\bibfnamefont {S.~K.}\ \bibnamefont {Baek}},\ }\href@noop
  {} {\bibfield  {journal} {\bibinfo  {journal} {J. Korean Phys. Soc,}\
  }\textbf {\bibinfo {volume} {66}},\ \bibinfo {pages} {739} (\bibinfo {year}
  {2015})}\BibitemShut {NoStop}%
\bibitem [{\citenamefont {Jauch}\ and\ \citenamefont {Baron}(1972)}]{jauch}%
  \BibitemOpen
  \bibfield  {author} {\bibinfo {author} {\bibfnamefont {J.~M.}\ \bibnamefont
  {Jauch}}\ and\ \bibinfo {author} {\bibfnamefont {J.~G.}\ \bibnamefont
  {Baron}},\ }\href@noop {} {\bibfield  {journal} {\bibinfo  {journal} {Helv.
  Phys. Acta}\ }\textbf {\bibinfo {volume} {45}},\ \bibinfo {pages} {220}
  (\bibinfo {year} {1972})}\BibitemShut {NoStop}%
\bibitem [{\citenamefont {Messiah}(1961)}]{messiah}%
  \BibitemOpen
  \bibfield  {author} {\bibinfo {author} {\bibfnamefont {A.}~\bibnamefont
  {Messiah}},\ }\href@noop {} {\emph {\bibinfo {title} {Quantum mechanics}}}\
  (\bibinfo  {publisher} {North-Holland Pub. Co.},\ \bibinfo {address}
  {Amsterdam},\ \bibinfo {year} {1961})\BibitemShut {NoStop}%
\bibitem [{\citenamefont {Bender}\ \emph {et~al.}(2005)\citenamefont {Bender},
  \citenamefont {Brody},\ and\ \citenamefont {Meister}}]{bender}%
  \BibitemOpen
  \bibfield  {author} {\bibinfo {author} {\bibfnamefont {C.~M.}\ \bibnamefont
  {Bender}}, \bibinfo {author} {\bibfnamefont {D.~C.}\ \bibnamefont {Brody}}, \
  and\ \bibinfo {author} {\bibfnamefont {B.~K.}\ \bibnamefont {Meister}},\
  }\href@noop {} {\bibfield  {journal} {\bibinfo  {journal} {Proc. R. Soc. A}\
  }\textbf {\bibinfo {volume} {461}},\ \bibinfo {pages} {733} (\bibinfo {year}
  {2005})}\BibitemShut {NoStop}%
\bibitem [{\citenamefont {Kim}\ and\ \citenamefont {Kim}(2011)}]{kim2}%
  \BibitemOpen
  \bibfield  {author} {\bibinfo {author} {\bibfnamefont {K.-H.}\ \bibnamefont
  {Kim}}\ and\ \bibinfo {author} {\bibfnamefont {S.~W.}\ \bibnamefont {Kim}},\
  }\href@noop {} {\bibfield  {journal} {\bibinfo  {journal} {Phys. Rev. E}\
  }\textbf {\bibinfo {volume} {84}},\ \bibinfo {pages} {012101} (\bibinfo
  {year} {2011})}\BibitemShut {NoStop}%
\bibitem [{\citenamefont {{Lewis~Jr.}}(1968)}]{lewis}%
  \BibitemOpen
  \bibfield  {author} {\bibinfo {author} {\bibfnamefont {H.~R.}\ \bibnamefont
  {{Lewis~Jr.}}},\ }\href@noop {} {\bibfield  {journal} {\bibinfo  {journal}
  {J. Math. Phys.}\ }\textbf {\bibinfo {volume} {9}},\ \bibinfo {pages} {1976}
  (\bibinfo {year} {1968})}\BibitemShut {NoStop}%
\bibitem [{\citenamefont {Ji}\ \emph {et~al.}(1995)\citenamefont {Ji},
  \citenamefont {Kim},\ and\ \citenamefont {Kim}}]{ji}%
  \BibitemOpen
  \bibfield  {author} {\bibinfo {author} {\bibfnamefont {J.-Y.}\ \bibnamefont
  {Ji}}, \bibinfo {author} {\bibfnamefont {J.~K.}\ \bibnamefont {Kim}}, \ and\
  \bibinfo {author} {\bibfnamefont {S.~P.}\ \bibnamefont {Kim}},\ }\href@noop
  {} {\bibfield  {journal} {\bibinfo  {journal} {Phys. Rev. A}\ }\textbf
  {\bibinfo {volume} {51}},\ \bibinfo {pages} {4268} (\bibinfo {year}
  {1995})}\BibitemShut {NoStop}%
\bibitem [{\citenamefont {Truscott}(1993)}]{truscott}%
  \BibitemOpen
  \bibfield  {author} {\bibinfo {author} {\bibfnamefont {W.~S.}\ \bibnamefont
  {Truscott}},\ }\href@noop {} {\bibfield  {journal} {\bibinfo  {journal}
  {Phys. Rev. Lett.}\ }\textbf {\bibinfo {volume} {70}},\ \bibinfo {pages}
  {1900} (\bibinfo {year} {1993})}\BibitemShut {NoStop}%
\bibitem [{\citenamefont {Feng}(2001)}]{feng}%
  \BibitemOpen
  \bibfield  {author} {\bibinfo {author} {\bibfnamefont {M.}~\bibnamefont
  {Feng}},\ }\href@noop {} {\bibfield  {journal} {\bibinfo  {journal} {Phys.
  Rev. A}\ }\textbf {\bibinfo {volume} {64}},\ \bibinfo {pages} {034101}
  (\bibinfo {year} {2001})}\BibitemShut {NoStop}%
\bibitem [{\citenamefont {Park}(2002)}]{park02}%
  \BibitemOpen
  \bibfield  {author} {\bibinfo {author} {\bibfnamefont {T.~J.}\ \bibnamefont
  {Park}},\ }\href@noop {} {\bibfield  {journal} {\bibinfo  {journal} {Bull.
  Korean Chem. Soc.}\ }\textbf {\bibinfo {volume} {23}},\ \bibinfo {pages}
  {355} (\bibinfo {year} {2002})}\BibitemShut {NoStop}%
\bibitem [{\citenamefont {Pezz\'{e}}\ \emph {et~al.}(2005)\citenamefont
  {Pezz\'{e}}, \citenamefont {Smerzi}, \citenamefont {Berman}, \citenamefont
  {Bishop},\ and\ \citenamefont {Collins}}]{pezze}%
  \BibitemOpen
  \bibfield  {author} {\bibinfo {author} {\bibfnamefont {L.}~\bibnamefont
  {Pezz\'{e}}}, \bibinfo {author} {\bibfnamefont {A.}~\bibnamefont {Smerzi}},
  \bibinfo {author} {\bibfnamefont {G.~P.}\ \bibnamefont {Berman}}, \bibinfo
  {author} {\bibfnamefont {A.~R.}\ \bibnamefont {Bishop}}, \ and\ \bibinfo
  {author} {\bibfnamefont {L.~A.}\ \bibnamefont {Collins}},\ }\href@noop {}
  {\bibfield  {journal} {\bibinfo  {journal} {New J. Phys.}\ }\textbf {\bibinfo
  {volume} {7}},\ \bibinfo {pages} {85} (\bibinfo {year} {2005})}\BibitemShut
  {NoStop}%
\bibitem [{\citenamefont {Hohenester}\ \emph {et~al.}(2007)\citenamefont
  {Hohenester}, \citenamefont {Rekdal}, \citenamefont {Borz\`{i}},\ and\
  \citenamefont {Schmiedmayer}}]{hohenester}%
  \BibitemOpen
  \bibfield  {author} {\bibinfo {author} {\bibfnamefont {U.}~\bibnamefont
  {Hohenester}}, \bibinfo {author} {\bibfnamefont {P.~K.}\ \bibnamefont
  {Rekdal}}, \bibinfo {author} {\bibfnamefont {A.}~\bibnamefont {Borz\`{i}}}, \
  and\ \bibinfo {author} {\bibfnamefont {J.}~\bibnamefont {Schmiedmayer}},\
  }\href@noop {} {\bibfield  {journal} {\bibinfo  {journal} {Phys. Rev. A}\
  }\textbf {\bibinfo {volume} {75}},\ \bibinfo {pages} {023602} (\bibinfo
  {year} {2007})}\BibitemShut {NoStop}%
\bibitem [{\citenamefont {Grond}\ \emph
  {et~al.}(2009{\natexlab{a}})\citenamefont {Grond}, \citenamefont
  {Schmiedmayer},\ and\ \citenamefont {Hohenester}}]{grond09a}%
  \BibitemOpen
  \bibfield  {author} {\bibinfo {author} {\bibfnamefont {J.}~\bibnamefont
  {Grond}}, \bibinfo {author} {\bibfnamefont {J.}~\bibnamefont {Schmiedmayer}},
  \ and\ \bibinfo {author} {\bibfnamefont {U.}~\bibnamefont {Hohenester}},\
  }\href@noop {} {\bibfield  {journal} {\bibinfo  {journal} {Phys. Rev. A}\
  }\textbf {\bibinfo {volume} {79}},\ \bibinfo {pages} {021603(R)} (\bibinfo
  {year} {2009}{\natexlab{a}})}\BibitemShut {NoStop}%
\bibitem [{\citenamefont {Grond}\ \emph
  {et~al.}(2009{\natexlab{b}})\citenamefont {Grond}, \citenamefont
  {{von~Winckel}}, \citenamefont {Schmiedmayer},\ and\ \citenamefont
  {Hohenester}}]{grond09b}%
  \BibitemOpen
  \bibfield  {author} {\bibinfo {author} {\bibfnamefont {J.}~\bibnamefont
  {Grond}}, \bibinfo {author} {\bibfnamefont {G.}~\bibnamefont
  {{von~Winckel}}}, \bibinfo {author} {\bibfnamefont {J.}~\bibnamefont
  {Schmiedmayer}}, \ and\ \bibinfo {author} {\bibfnamefont {U.}~\bibnamefont
  {Hohenester}},\ }\href@noop {} {\bibfield  {journal} {\bibinfo  {journal}
  {Phys. Rev. A}\ }\textbf {\bibinfo {volume} {80}},\ \bibinfo {pages} {053625}
  (\bibinfo {year} {2009}{\natexlab{b}})}\BibitemShut {NoStop}%
\bibitem [{\citenamefont {Masuda}\ and\ \citenamefont
  {Nakamura}(2008)}]{masuda08}%
  \BibitemOpen
  \bibfield  {author} {\bibinfo {author} {\bibfnamefont {S.}~\bibnamefont
  {Masuda}}\ and\ \bibinfo {author} {\bibfnamefont {K.}~\bibnamefont
  {Nakamura}},\ }\href@noop {} {\bibfield  {journal} {\bibinfo  {journal}
  {Phys. Rev. A}\ }\textbf {\bibinfo {volume} {78}},\ \bibinfo {pages} {062108}
  (\bibinfo {year} {2008})}\BibitemShut {NoStop}%
\bibitem [{\citenamefont {Masuda}\ and\ \citenamefont
  {Nakamura}(2009)}]{masuda09}%
  \BibitemOpen
  \bibfield  {author} {\bibinfo {author} {\bibfnamefont {S.}~\bibnamefont
  {Masuda}}\ and\ \bibinfo {author} {\bibfnamefont {K.}~\bibnamefont
  {Nakamura}},\ }\href@noop {} {\bibfield  {journal} {\bibinfo  {journal}
  {Proc. R. Soc. A}\ }\textbf {\bibinfo {volume} {466}},\ \bibinfo {pages}
  {1135} (\bibinfo {year} {2009})}\BibitemShut {NoStop}%
\bibitem [{\citenamefont {Torrontegui}\ \emph {et~al.}(2013)\citenamefont
  {Torrontegui}, \citenamefont {Mart\'{i}nez-Garaot}, \citenamefont {Modugno},
  \citenamefont {Chen},\ and\ \citenamefont {Muga}}]{torrontegui}%
  \BibitemOpen
  \bibfield  {author} {\bibinfo {author} {\bibfnamefont {E.}~\bibnamefont
  {Torrontegui}}, \bibinfo {author} {\bibfnamefont {S.}~\bibnamefont
  {Mart\'{i}nez-Garaot}}, \bibinfo {author} {\bibfnamefont {M.}~\bibnamefont
  {Modugno}}, \bibinfo {author} {\bibfnamefont {X.}~\bibnamefont {Chen}}, \
  and\ \bibinfo {author} {\bibfnamefont {J.~G.}\ \bibnamefont {Muga}},\
  }\href@noop {} {\bibfield  {journal} {\bibinfo  {journal} {Phys. Rev. A}\
  }\textbf {\bibinfo {volume} {87}},\ \bibinfo {pages} {033630} (\bibinfo
  {year} {2013})}\BibitemShut {NoStop}%
\bibitem [{\citenamefont {Shin}\ \emph {et~al.}(2004)\citenamefont {Shin},
  \citenamefont {Saba}, \citenamefont {Pasquini}, \citenamefont {Ketterle},
  \citenamefont {Pritchard},\ and\ \citenamefont {Leanhardt}}]{shin}%
  \BibitemOpen
  \bibfield  {author} {\bibinfo {author} {\bibfnamefont {Y.}~\bibnamefont
  {Shin}}, \bibinfo {author} {\bibfnamefont {M.}~\bibnamefont {Saba}}, \bibinfo
  {author} {\bibfnamefont {T.~A.}\ \bibnamefont {Pasquini}}, \bibinfo {author}
  {\bibfnamefont {W.}~\bibnamefont {Ketterle}}, \bibinfo {author}
  {\bibfnamefont {D.~E.}\ \bibnamefont {Pritchard}}, \ and\ \bibinfo {author}
  {\bibfnamefont {A.~E.}\ \bibnamefont {Leanhardt}},\ }\href@noop {} {\bibfield
   {journal} {\bibinfo  {journal} {Phys. Rev. Lett.}\ }\textbf {\bibinfo
  {volume} {92}},\ \bibinfo {pages} {050405} (\bibinfo {year}
  {2004})}\BibitemShut {NoStop}%
\bibitem [{\citenamefont {Albiez}\ \emph {et~al.}(2005)\citenamefont {Albiez},
  \citenamefont {Gati}, \citenamefont {F{\"o}lling}, \citenamefont {Hunsmann},
  \citenamefont {Cristiani},\ and\ \citenamefont {Oberthaler}}]{albiez}%
  \BibitemOpen
  \bibfield  {author} {\bibinfo {author} {\bibfnamefont {M.}~\bibnamefont
  {Albiez}}, \bibinfo {author} {\bibfnamefont {R.}~\bibnamefont {Gati}},
  \bibinfo {author} {\bibfnamefont {J.}~\bibnamefont {F{\"o}lling}}, \bibinfo
  {author} {\bibfnamefont {S.}~\bibnamefont {Hunsmann}}, \bibinfo {author}
  {\bibfnamefont {M.}~\bibnamefont {Cristiani}}, \ and\ \bibinfo {author}
  {\bibfnamefont {M.~K.}\ \bibnamefont {Oberthaler}},\ }\href@noop {}
  {\bibfield  {journal} {\bibinfo  {journal} {Phys. Rev. Lett.}\ }\textbf
  {\bibinfo {volume} {95}},\ \bibinfo {pages} {010402} (\bibinfo {year}
  {2005})}\BibitemShut {NoStop}%
\bibitem [{\citenamefont {{Van~den~Broeck}}(2005)}]{broeck}%
  \BibitemOpen
  \bibfield  {author} {\bibinfo {author} {\bibfnamefont {C.}~\bibnamefont
  {{Van~den~Broeck}}},\ }\href@noop {} {\bibfield  {journal} {\bibinfo
  {journal} {Phys. Rev. Lett.}\ }\textbf {\bibinfo {volume} {95}},\ \bibinfo
  {pages} {190602} (\bibinfo {year} {2005})}\BibitemShut {NoStop}%
\bibitem [{\citenamefont {Esposito}\ \emph {et~al.}(2009)\citenamefont
  {Esposito}, \citenamefont {Lindenberg},\ and\ \citenamefont
  {{Van~den~Broeck}}}]{esposito}%
  \BibitemOpen
  \bibfield  {author} {\bibinfo {author} {\bibfnamefont {M.}~\bibnamefont
  {Esposito}}, \bibinfo {author} {\bibfnamefont {K.}~\bibnamefont
  {Lindenberg}}, \ and\ \bibinfo {author} {\bibfnamefont {C.}~\bibnamefont
  {{Van~den~Broeck}}},\ }\href@noop {} {\bibfield  {journal} {\bibinfo
  {journal} {Phys. Rev. Lett.}\ }\textbf {\bibinfo {volume} {102}},\ \bibinfo
  {pages} {130602} (\bibinfo {year} {2009})}\BibitemShut {NoStop}%
\bibitem [{\citenamefont {Esposito}\ \emph {et~al.}(2010)\citenamefont
  {Esposito}, \citenamefont {Kawai}, \citenamefont {Lindenberg},\ and\
  \citenamefont {{Van~den~Broeck}}}]{kawai}%
  \BibitemOpen
  \bibfield  {author} {\bibinfo {author} {\bibfnamefont {M.}~\bibnamefont
  {Esposito}}, \bibinfo {author} {\bibfnamefont {R.}~\bibnamefont {Kawai}},
  \bibinfo {author} {\bibfnamefont {K.}~\bibnamefont {Lindenberg}}, \ and\
  \bibinfo {author} {\bibfnamefont {C.}~\bibnamefont {{Van~den~Broeck}}},\
  }\href@noop {} {\bibfield  {journal} {\bibinfo  {journal} {Phys. Rev. Lett.}\
  }\textbf {\bibinfo {volume} {105}},\ \bibinfo {pages} {150603} (\bibinfo
  {year} {2010})}\BibitemShut {NoStop}%
\bibitem [{\citenamefont {Campbell}(2009)}]{campbell}%
  \BibitemOpen
  \bibfield  {author} {\bibinfo {author} {\bibfnamefont {J.}~\bibnamefont
  {Campbell}},\ }\href@noop {} {\bibfield  {journal} {\bibinfo  {journal} {J.
  Phys. A}\ }\textbf {\bibinfo {volume} {42}},\ \bibinfo {pages} {365212}
  (\bibinfo {year} {2009})}\BibitemShut {NoStop}%
\bibitem [{\citenamefont {Boas}(2006)}]{boas}%
  \BibitemOpen
  \bibfield  {author} {\bibinfo {author} {\bibfnamefont {M.~L.}\ \bibnamefont
  {Boas}},\ }\href@noop {} {\emph {\bibinfo {title} {Mathematical Methods in
  the Physical Sciences}}},\ \bibinfo {edition} {3rd}\ ed.\ (\bibinfo
  {publisher} {Wiley},\ \bibinfo {address} {Hoboken, NJ},\ \bibinfo {year}
  {2006})\BibitemShut {NoStop}%
\bibitem [{\citenamefont {Newman}(2013)}]{newman}%
  \BibitemOpen
  \bibfield  {author} {\bibinfo {author} {\bibfnamefont {M.~E.~J.}\
  \bibnamefont {Newman}},\ }\href@noop {} {\emph {\bibinfo {title}
  {Computational Physics}}}\ (\bibinfo  {publisher} {CreateSpace Independent},\
  \bibinfo {address} {United States},\ \bibinfo {year} {2013})\BibitemShut
  {NoStop}%
\bibitem [{\citenamefont {Gea-Banacloche}(2002)}]{gea}%
  \BibitemOpen
  \bibfield  {author} {\bibinfo {author} {\bibfnamefont {J.}~\bibnamefont
  {Gea-Banacloche}},\ }\href@noop {} {\bibfield  {journal} {\bibinfo  {journal}
  {Am. J. Phys.}\ }\textbf {\bibinfo {volume} {70}},\ \bibinfo {pages} {307}
  (\bibinfo {year} {2002})}\BibitemShut {NoStop}%
\bibitem [{\citenamefont {Lakner}\ and\ \citenamefont
  {Peternelj}(2003)}]{lakner}%
  \BibitemOpen
  \bibfield  {author} {\bibinfo {author} {\bibfnamefont {M.}~\bibnamefont
  {Lakner}}\ and\ \bibinfo {author} {\bibfnamefont {J.}~\bibnamefont
  {Peternelj}},\ }\href@noop {} {\bibfield  {journal} {\bibinfo  {journal} {Am.
  J. Phys.}\ }\textbf {\bibinfo {volume} {71}},\ \bibinfo {pages} {519}
  (\bibinfo {year} {2003})}\BibitemShut {NoStop}%
\bibitem [{\citenamefont {van Enk}(2009)}]{enk}%
  \BibitemOpen
  \bibfield  {author} {\bibinfo {author} {\bibfnamefont {S.~J.}\ \bibnamefont
  {van Enk}},\ }\href@noop {} {\bibfield  {journal} {\bibinfo  {journal} {Am.
  J. Phys.}\ }\textbf {\bibinfo {volume} {77}},\ \bibinfo {pages} {140}
  (\bibinfo {year} {2009})}\BibitemShut {NoStop}%
\bibitem [{\citenamefont {Kim}\ and\ \citenamefont {Kim}(2012)}]{khkim}%
  \BibitemOpen
  \bibfield  {author} {\bibinfo {author} {\bibfnamefont {K.-H.}\ \bibnamefont
  {Kim}}\ and\ \bibinfo {author} {\bibfnamefont {S.~W.}\ \bibnamefont {Kim}},\
  }\href@noop {} {\bibfield  {journal} {\bibinfo  {journal} {J. Korean Phys.
  Soc.}\ }\textbf {\bibinfo {volume} {61}},\ \bibinfo {pages} {1187} (\bibinfo
  {year} {2012})}\BibitemShut {NoStop}%
\bibitem [{\citenamefont {Meyrath}\ \emph {et~al.}(2005)\citenamefont
  {Meyrath}, \citenamefont {Schreck}, \citenamefont {Hanssen}, \citenamefont
  {Chuu},\ and\ \citenamefont {Raizen}}]{meyrath}%
  \BibitemOpen
  \bibfield  {author} {\bibinfo {author} {\bibfnamefont {T.~P.}\ \bibnamefont
  {Meyrath}}, \bibinfo {author} {\bibfnamefont {F.}~\bibnamefont {Schreck}},
  \bibinfo {author} {\bibfnamefont {J.~L.}\ \bibnamefont {Hanssen}}, \bibinfo
  {author} {\bibfnamefont {C.-S.}\ \bibnamefont {Chuu}}, \ and\ \bibinfo
  {author} {\bibfnamefont {M.~G.}\ \bibnamefont {Raizen}},\ }\href@noop {}
  {\bibfield  {journal} {\bibinfo  {journal} {Phys. Rev. A}\ }\textbf {\bibinfo
  {volume} {71}},\ \bibinfo {pages} {041604(R)} (\bibinfo {year}
  {2005})}\BibitemShut {NoStop}%
\bibitem [{\citenamefont {Henderson}\ \emph {et~al.}(2006)\citenamefont
  {Henderson}, \citenamefont {Kelkar}, \citenamefont {Li}, \citenamefont
  {Guti{\'e}rrez-Medina},\ and\ \citenamefont {Raizen}}]{henderson}%
  \BibitemOpen
  \bibfield  {author} {\bibinfo {author} {\bibfnamefont {K.}~\bibnamefont
  {Henderson}}, \bibinfo {author} {\bibfnamefont {H.}~\bibnamefont {Kelkar}},
  \bibinfo {author} {\bibfnamefont {T.~C.}\ \bibnamefont {Li}}, \bibinfo
  {author} {\bibfnamefont {B.}~\bibnamefont {Guti{\'e}rrez-Medina}}, \ and\
  \bibinfo {author} {\bibfnamefont {M.~G.}\ \bibnamefont {Raizen}},\
  }\href@noop {} {\bibfield  {journal} {\bibinfo  {journal} {EPL}\ }\textbf
  {\bibinfo {volume} {75}},\ \bibinfo {pages} {392} (\bibinfo {year}
  {2006})}\BibitemShut {NoStop}%
\bibitem [{\citenamefont {Zhang}\ \emph {et~al.}(2008)\citenamefont {Zhang},
  \citenamefont {Nho},\ and\ \citenamefont {Landau}}]{zhang}%
  \BibitemOpen
  \bibfield  {author} {\bibinfo {author} {\bibfnamefont {C.}~\bibnamefont
  {Zhang}}, \bibinfo {author} {\bibfnamefont {K.}~\bibnamefont {Nho}}, \ and\
  \bibinfo {author} {\bibfnamefont {D.~P.}\ \bibnamefont {Landau}},\
  }\href@noop {} {\bibfield  {journal} {\bibinfo  {journal} {Phys. Rev. A}\
  }\textbf {\bibinfo {volume} {77}},\ \bibinfo {pages} {025601} (\bibinfo
  {year} {2008})}\BibitemShut {NoStop}%
\bibitem [{\citenamefont {Antunes}\ \emph {et~al.}(2006)\citenamefont
  {Antunes}, \citenamefont {Lombardo}, \citenamefont {Monteoliva},\ and\
  \citenamefont {Villar}}]{tunnel}%
  \BibitemOpen
  \bibfield  {author} {\bibinfo {author} {\bibfnamefont {N.~D.}\ \bibnamefont
  {Antunes}}, \bibinfo {author} {\bibfnamefont {F.~C.}\ \bibnamefont
  {Lombardo}}, \bibinfo {author} {\bibfnamefont {D.}~\bibnamefont
  {Monteoliva}}, \ and\ \bibinfo {author} {\bibfnamefont {P.~I.}\ \bibnamefont
  {Villar}},\ }\href@noop {} {\bibfield  {journal} {\bibinfo  {journal} {Phys.
  Rev. E}\ }\textbf {\bibinfo {volume} {73}},\ \bibinfo {pages} {066105}
  (\bibinfo {year} {2006})}\BibitemShut {NoStop}%
\end{thebibliography}
%
\end{document}